\documentclass[utf8]{frontiersFPHY} 

\usepackage{url,hyperref,lineno,microtype,subcaption,ulem}
\usepackage[onehalfspacing]{setspace}

\renewcommand{\vec}{\mathbf}

\def\keyFont{\fontsize{8}{11}\helveticabold }
\def\firstAuthorLast{Moreno {et~al.}} 
\def\Authors{Eduardo Moreno\,$^{1}$, Robert Gro{\ss}mann\,$^{2}$, Carsten Beta\,$^{2}$ and Sergio Alonso\,$^{1,*}$}
\def\Address{$^{1}$ Department of Physics, Universitat Politècnica de Catalunya, Barcelona, Spain \\
$^{2}$ Institute of Physics and Astronomy, University of Potsdam, Potsdam, Germany   
}
\def\corrAuthor{Sergio Alonso}
\def\corrEmail{s.alonso@upc.edu}

\begin{document}
\onecolumn
\firstpage{1}

\title[A computational study of interacting amoeboid cells]{From single to collective motion of social amoebae: a computational study of interacting cells} 

\author[\firstAuthorLast ]{\Authors} 
\address{} 
\correspondance{} 

\extraAuth{}

\maketitle

\begin{abstract}

\noindent
The coupling of the internal mechanisms of cell polarization to cell shape deformations and subsequent cell crawling poses many interdisciplinary scientific challenges.
Several mathematical approaches have been proposed to model the coupling of both processes, where one of the most successful methods relies on a phase field that encodes the morphology of the cell, together with the integration of partial differential equations that account for the polarization mechanism inside the cell domain as defined by the phase field.
This approach has been previously employed to model the motion of single cells of the social amoeba \textit{Dictyostelium~discoideum}, a widely used model organism to study actin-driven motility and chemotaxis of eukaryotic cells.
Besides single cell motility, \textit{Dictyostelium~discoideum} is also well-known for its collective behavior.
Here, we extend the previously introduced model for single cell motility to describe the collective motion of large populations of interacting amoebae by including repulsive interactions between the cells.
We performed numerical simulations of this model, first characterizing the motion of single cells in terms of their polarity and velocity vectors.
We then systematically studied the collisions between two cells that provided the basic interaction scenarios also observed in larger ensembles of interacting amoebae.
Finally, the relevance of the cell density was analyzed, revealing a systematic decrease of the motility with density, associated with the formation of transient cell clusters that emerge in this system even though our model does not include any attractive interactions between cells.
%
This model is a prototypical active matter system for the investigation of the emergent collective dynamics of deformable, self-driven cells with a highly complex, nonlinear coupling of cell shape deformations, self-propulsion and repulsive cell-cell interactions.
Understanding these self-organization processes of cells like their autonomous aggregation is of high relevance as collective amoeboid motility is part of wound healing, embryonic morphogenesis or pathological processes like the spreading of metastatic cancer cells.

\tiny
 \keyFont{ \section{Keywords:} cell motility, cell polarity, reaction-diffusion models, cell-cell interactions, phase field model, collective motion, active matter} 
\end{abstract}

\section{Introduction}
Collective migration has been extensively studied in a wide range of different systems, such as bird flocks, fish schools, or crowds of pedestrians~\cite{vicsek2021collective}.
However, not only at the level of higher organisms but also at the cellular scale, collective migration is key to many essential biological processes~\citep{friedl2009collective}. 
During wound healing, for example, groups of cells are moving towards the site of injury to remove bacterial infections and to close the tissue~\citep{poujade2007collective}.
Similarly, in the course of embryonic morphogenesis~\citep{heisenberg2013forces}, cells collectively migrate to distinct locations in the embryo to fulfil specific tasks~\citep{dormann2003chemotactic,montell2003border}.
Collective cell migration is also involved in pathological processes, such as the spreading of metastatic cancer cells~\citep{cheung2016collective, friedl2012classifying}, where groups of cells migrate through tissue and initiate the formation of tumors.

Many properties of the movement of single adherent cells on flat substrates are well understood.
In particular, the dynamics of the actin cortex that relies on the continuous polymerization and depolymerization of actin filaments has been intensely investigated \cite{blanchoin2014actin}. 
The actin system is known to drive the formation of lamellipodia and protrusions that push the cell forward~\citep{mitchison1996actin,alberts1989molecular}. 
In contrast to the motility of single cells, much less is known about the mechanisms by which groups of cells organize and coordinate their movement.
Recent advances in microscopy imaging techniques combined with cell tracking methods permit us to study collective migration patterns of large numbers of cells over extended periods of time with single cell resolution.
This provides a better understanding of how groups of cells coordinate their movement and may allow to develop, for example, therapeutic drugs. 
Note that collective effects in populations of motile cells do not simply arise as a consequence of their free random movement. 
Instead, when moving in groups, cells may act in a coordinated way, resulting in collective modes of locomotion that clearly differ from the way they move as isolated individuals.
Well-known examples arise when cells form a dense monolayer~\citep{farhadifar2007influence} or when groups of cells are governed by a few leader cells~\citep{kabla2012collective}.
However, as long as they are not too tightly packed, they typically continue moving as individuals and occasionally form clusters that show coordinated patterns and localized bursts of velocities~\citep{petitjean2010velocity}. 

Together with progress in the experimental studies of collective cell migration, also efforts in physical modeling of this process have increased.
Describing the motion of cells in a quantitative manner requires solving equations that incorporate intracellular processes as well as interactions with the environment; for a review of different physical modeling approaches, see Refs.~\citep{aranson2016physical,alert2020physical,camley2017physical} and references therein.
A simplistic, well-known approach to describe collective motion from a physical point of view is the Vicsek model~\citep{vicsek1995novel}, where active agents are described as point-like, self-propelled particles interacting via heuristic velocity-alignment rules of different symmetry, which effectively mimic the effects of collisions~\cite{gregoire2004onset,bar2020self}. 
In order to model cell-cell interactions in a more realistic fashion, however, it is necessary to take the spatial extension of cells and their dynamic shape changes into account, which cannot be achieved by a point particle description.
In general, particle shape is known to crucially affect the emergent patterns in ensembles of active particles~\cite{grossmann2020particle}:~whereas self-propelled discs undergo aggregation and motility-induced phase separation, even in the absence of attractive interactions~\cite{buttinoni2013dynamical,caes2015motility}, self-propelled rod-shaped particles were shown to display a variety of self-organized, complex patterns of different symmetry such as polar clusters, nematic bands or mesoscale turbulence~\cite{bar2020self}. 
The pattern formation in soft deformable self-propelled particles that can adopt different elliptical shapes depending on the strength of self-propulsion is correspondingly complex and rich~\cite{menzel2012soft}; in the latter work, however, particle shape was not considered to be dictated by an intracellular biochemical reaction but heuristically introduced via a direct coupling to the strength of self-propulsion. 
More realistic cell shapes were successfully incorporated using the lattice based Cellular Potts Model~\citep{glazier1993simulation,graner1992simulation}.
Here, the cell is represented by a group of connected pixels of a given size, where every pixel is regularly updated according to certain probabilistic rules.
A major limitation of this method is that the dynamics of pixel updates is artificial and difficult to relate to the real-time dynamics of cell motion.
Alternatively, also active network models, such as Vertex~\citep{nagai2001dynamic,staple2010mechanics} and Voronoi models~\citep{li2014coherent,bi2016motility}, are used to study cell migration, describing tissues as a network of polygonal cells~\citep{alt2017vertex}.
Here, however, a limitation is that these models take neither internal processes in the cells nor anisotropic active stresses in the tissue into account.
We are interested in mathematical models which are able to describe the emergence of collective dynamics together with the intracellular pattern formation responsible for the polarization of single cells. 
An approach that became increasingly popular in recent years to model cell migration is the phase field method that readily incorporates evolving geometries without the need of explicitly tracking the domain boundaries. 
In cell motility modeling, the phase field defines the cell shape and position, being one inside and zero outside the cell, and it maintains the correct boundary conditions while the border is moving~\citep{kockelkoren2003computational,camley2013periodic}.
The phase field approach has been used to study the persistent~\citep{shao2010computational,ziebert2012model,shao2012coupling}, bipedal~\citep{lober2014modeling}, and rotary motion~\citep{camley2017crawling} in keratocyte-like cells.
More recently, it was also employed to model amoeboid movement of cells such as neutrophils~\citep{najem2013phase} and {\it Dictyostelium discoideum}~\citep{alonso2018modeling}.
In addition, other features, such as viscoelasticity~\citep{moure2016computational,moure2018three}, the effect of intracellular biochemical waves~\citep{taniguchi2013phase}, and wave-driven cytofission of cells~\citep{flemming2020cortical} have been successfully modeled based on the phase field approach.
There is also work related to the use of phase fields to model the collective motion of eukaryotic cells:~in early studies, shape changes due to interactions~\citep{coburn2013tactile} and the rearrangement of cells in clusters~\citep{nonomura2012study} were considered.
Later, monolayers of deformable motile cells where investigated, revealing a melting transition in these systems~\citep{palmieri2015multiple,peyret2019sustained,mueller2019emergence,zhang2020active,loewe2020solid}.
Also the impact of collisions between deformable cells has been studied within the phase field approach, focusing on the resulting alignment and collective motion of cells~\citep{camley2014polarity,lober2015collisions,kulawiak2016modeling}.




In this article, we focus on the collective motion of amoeboid cells, motivated by the relevance of amoeboid motility for immune responses and cancer spreading.
Also the social amoeba {\it Dictyostelium discoideum}~(\textit{D.~discoideum}), a well-established model organism to study amoeboid locomotion, is known for its collective behavior \cite{gregor2010onset}.
In the course of its starvation-induced life cycle, {\it D.~discoideum} cells organize into large multicellular colonies, a process mediated by self-organized waves of chemoattractant that travel across spatially distributed populations of cells to guide their aggregation process~\cite{siegert1989digital}.
The underlying set of interactions is complex and several models have focused on the mechanisms of periodic chemoattractant signaling and the resulting emergent waves \cite{grace2015regulation,vidal2019spontaneous}. 
Here, we will not take the specific aspects of the signaling mechanisms of {\it D.~discoideum} into account.
Instead, we will concentrate on generic aspects of collective motion that may arise from pure mechanical interactions of amoeboid cells.
All forms of mutual chemical signaling between the cells are excluded.
For our numerical study, we will rely on a mathematical model based on a reaction-diffusion system that is embedded into a phase field, accounting for the intracellular dynamics and driving the evolution of the cell contour.
This model has been previously established to reproduce the dynamics of single {\it D.~discoideum} cells under different conditions~\citep{alonso2018modeling,moreno2020modeling} and will now be extended to study ensembles of interacting cells.
In particular, the ability to describe deformable objects with a phase field representation will be exploited to study the mechanical interaction between these motile, deformable particles that can be seen as a simplistic representation of ensembles of migrating amoeboid cells.
First, we will study the interactions between two cells and later increase the density of cells to investigate the dynamics of larger populations.

%
In summary, we developed a mathematical model that enables us to study the dynamics of deformable, actively moving entities, which can be applied to the interaction among cells in different physiological conditions and, in particular, to the motion of \textit{D.~discoideum}. We have obtained the density dependence of the average velocity and polarity of individual cells forming a swarm and describe the generation of transient clusters. 
This model is a prototypical active matter system for the investigation of the emergent collective dynamics of self-driven cells with a highly complex, nonlinear coupling of cell shape deformations, self-propulsion and repulsive cell-cell interactions.

\section{Methods and Model}

\subsection{Phase field description for the cell shape}
\label{sec:phasefield}

\noindent 
The computational model used here is an extended version of a previously introduced phase field model coupled to a stochastic bistable reaction-diffusion process~\citep{alonso2018modeling,flemming2020cortical}.
This model was originally designed to study the migratory behavior of an individual amoeba, such as a {\it D.~discoideum} cell, and is now extended to include interactions between several cells that are jointly simulated in the same computational domain.
The model is composed of a concentration variable~$c$ accounting for the biochemical processes occurring in the interior of the cell.
The variable $c$ represents a generic activatory signal that can be associated with upstream regulatory components of actin activity, such as activated Ras, PI3K or $\mbox{PIP}_3$.
The dynamics of $c$ is coupled to the evolution of the cell shape such that the cell boundary is protruding outward at locations of high values of $c$.
The cell shape is encoded by an auxiliary phase field $\phi$ which defines the area of the cell and varies from $\phi=1$ inside to $\phi=0$ outside of the cell.
In our simulations, each cell~(index~$i$) is represented by its own concentration variable~$c_i(\vec{x},t)$ and its own phase field $\phi_{i}(\vec{x},t)$.
Different cells interact via short-range repulsion that prevents overlap or adhesion.
The phase field evolves according to the following equation,
\begin{equation}\label{pf}
\tau \frac{\partial \phi_i}{\partial t} = \gamma \! \left(\nabla^2 \phi_i -\frac{G'(\phi_i)}{\epsilon^2}\right)
- \beta \! \left(\int \phi_i\, dA - A_0 \right)\left| \nabla \phi_i \right| + 
\alpha\, \phi_i\, c_i \left| \nabla \phi_i \right|  - m_{rep} \! \sum_{j=1,i\neq j}^{N} \phi_i \phi_j \left| \nabla \phi_i \right| , 
\end{equation}
where the first term on the right-hand side corresponds to the surface energy of the cell membrane related with a free energy which incorporates the derivative of a double well potential $G(\phi) = 18\,\phi^2\,(1-\phi)^2$.
The surface tension and the width of the cell boundary are parametrized by~$\gamma$ and~$\epsilon$, respectively. 
The second term enforces the cell area to stay close to a typical value~$A_0$, and the third term represents the active force that acts on the cell membrane, which is generated by the biochemical field $c_i$. These three terms can be derived from free energy arguments \cite{shao2010computational}.
The fourth term models cell-cell interactions controlled by parameter $m_{rep}$, thereby preventing cells from overlapping.
Here, $N$ corresponds to the total number of cells in the system. 
The parameters that control the surface tension, the strength of the area conservation constraint and the active force, $\gamma$, $\beta$ and $\alpha$, respectively, are kept constant in simulations, see Table~\ref{table1} for parameter values, taken to reproduce the dynamics of vegetative or starving \textit{D.~discoideum} cells~\cite{alonso2018modeling}.


\subsection{Reaction-diffusion equations for the intracellular biochemistry}

\noindent 
The dynamics of the biochemical component $c_i$ that diffuses and reacts inside the cell follows a noisy, bistable reaction kinetics,
\begin{equation}
\frac{\partial (\phi_i c_i) }{\partial t} = \nabla \! \left(\phi_i D \nabla  c_i\right) + \phi_i [k_a\, c_i\,(1-c_i)(c_i-\delta(c_i)) - \rho\, c_i] + \,\phi_i\,(1-\phi_i)\,\xi_i,  
\label{eqci}
\end{equation}
where $k_a$ is the reaction rate, $\rho$ the degradation rate, and $D$ the diffusivity of the component $c_i$.
The parameter $\delta$ controls the evolution of the concentration pattern and introduces a global feedback to maintain a fixed amount of $c_i$ inside the cell, thus imposing a mass-conservation condition in the model: 
\begin{equation}
\delta(c_i) = \delta_0  + M \left( \int c_i dA -C_0 \right)\! .
\label{eqdelta}
\end{equation}
The last term on the right-hand side of Eq.~\eqref{eqci} introduces noise, accounting for the stochastic nature of the reaction-diffusion processes occurring within the cell: the stochastic field~$\xi(\vec{x},t)$ follows an Ornstein-Uhlenbeck dynamics,
\begin{equation}\label{noise}
\frac{\partial \xi}{\partial t} = -k_{\eta}\, \xi + \eta\,,
\end{equation}
where $\eta(\vec{x},t)$ is a Gaussian white noise with zero mean, $\langle \eta \rangle=0$, and the variance $\langle \eta(\vec{x},t) \eta(\vec{x}\sp{\prime},t\sp{\prime})\rangle=2\sigma^2\delta(\vec{x}-\vec{x}\sp{\prime})\delta(t-t\sp{\prime})$. 
The relaxation rate~$k_{\eta}$ along with the reaction rate $k_a$ are key parameters in our model that control transitions between different forms of cell motility.

\subsection{Numerical methods}

\noindent 
Simulations were performed using finite differences with a spatial and temporal resolution of $\Delta x=0.15\,\mu m$ and $\Delta t=0.002\,s$, respectively.  We used the Euler-Maruyama method for the stochastic integration of the partial differential equations.
The total area considered for each cell was kept constant at $A_0=113$ $\mu m^2$ corresponding to a circular cell with radius $r=6\,\mu m$.
The area covered by the biochemical component $c_i$ was maintained at $C_0=28$ $\mu m^2$, corresponding to a quarter of the cell area.
The size of the grid and the number of cells were varied to explore different packing fractions $ N A_0 / L^2$ in the range from~$0.24$ to~$0.78$. 
Initial conditions of the intracellular concentration fields are chosen to produce polarized cells and promote collisions in the case of binary interaction scenarios, and isotropic in the case of multi-particle simulations. 

\subsection{Defining polarity and velocity vectors of migrating cells}
\label{sec:polarity}

\begin{figure}[h!]
\begin{center}
\includegraphics[width=\textwidth]{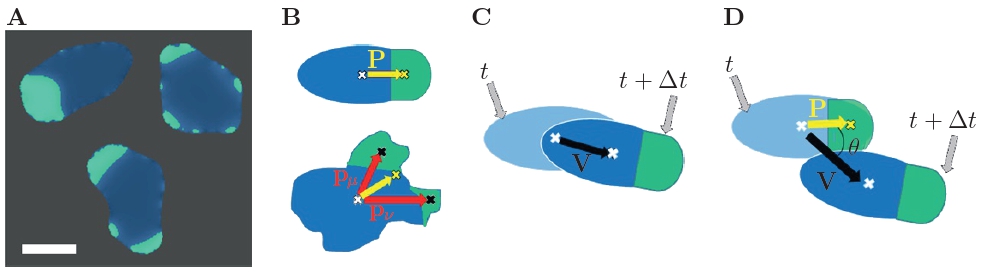}
\end{center}
\caption{\textbf{A}~Snapshots of the different cell phenotypes as observed in numerical simulations. Green patches indicate the presence of the biochemical component~$c$. Scale bar:~$10\mu m$. \textbf{B}~Illustration of the calculation of the cell polarity vector~$\vec{P}$ in the presence of one patch~(top) or several patches~(bottom). In the latter case, we consider the weighted averaged of the vectors~$\vec{p}_i$ shown in red. The weights are determined by the respective patch size. \textbf{C}~Illustration of the spatial displacement of a cell in the time interval from~$t$ to~$t + \Delta t$. \textbf{D}~Graphical representation of the angle~$\theta$ between cell polarity~$\vec{P}$ and velocity vector~$\vec{V}$.}\label{fig:1}
\end{figure}

\noindent 
Some characteristic cell morphologies as observed in numerical simulations are displayed in Figure~\ref{fig:1}. 
Patches of high concentration of the biochemical component $c_i$ that typically result in extensions/protrusions of the cell boundary are shown in green color in the respective images~(cf.~Figure~\ref{fig:1}A).
Similar to earlier experimental work, where fluorescently marked patches of PI3K were used to quantify the polarization of cells~\cite{park2016mechano}, we take advantage of the distribution of the biochemical field $c_i$ for that purpose here as follows. 
We measure the total area of the cell ($A_T$) and obtain the corresponding centroid coordinates ($x_T,y_T$).
We also measure the total area(s) of the patches covered by high values of the field $c_i$ ($A_{Ci}$) and obtain the centroid coordinates ($x_{C_i},y_{C_i}$) of these area(s). 
If multiple patches exist, we take the weighted average of the individual centroid coordinates to obtain the coordinate~($x_{CT},y_{CT}$). 
Eventually, we define a polarization vector $\vec{P}$ as the distance between the centroid coordinates ($x_T,y_T$) and the vector ($x_{CT},y_{CT}$).
Moreover, a velocity vector of cell propagation $\vec{V}$ is defined as the distance between the centroids of the cell at time $t$ and at subsequent time $t + \Delta t$ divided by the numerical time step $\Delta t$.
In addition, we furthermore calculate the angle $\theta$ between the two vectors $\vec{P}$ and $\vec{V}$.
In Figure~\ref{fig:1}B-D, the corresponding definitions of the coordinates, vectors, and the angle are shown.

\subsection{Classifying the collision scenarios between two cells}
\label{sec:scenarios}

\begin{figure}[h!]
\begin{center}
\includegraphics[width=\textwidth]{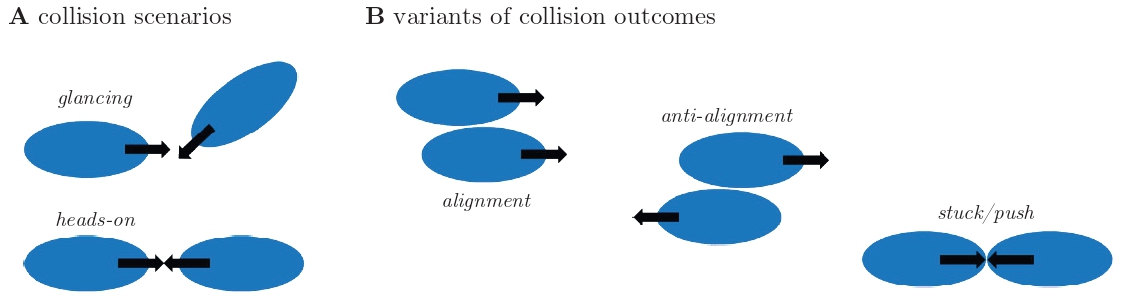}
\end{center}
\caption{\textbf{A}~Graphical illustration of the different collision scenarios and \textbf{B}~the respective outcome of the interaction. Initial conditions for binary collision are glancing~(top) and heads-on collision~(bottom). The interaction may result in alignment of cells, anti-alignment of velocities and cell polarities or cells may stuck/push heads-on, thereby impeding each others motion. }\label{fig:4}
\end{figure}

\noindent 
From all possible interactions between two cells that we can generate in numerical simulations, we focus on two specific types:~frontal and glancing collisions, see Figure~\ref{fig:4}A.
%
For both types, we observed three main interaction scenarios with different outcomes as illustrated in Figure~\ref{fig:4}.
In the first scenario, which we call \textit{alignment}, both cells face and migrate into the same direction after the collision.
In the second case, the \textit{anti-alignment} scenario, cells repel each other and move away in opposite directions after the collision.
Finally, upon collision, the motion of the two cells may also stall and both cells may remain on the spot, pushing head-on against each other for an extended period of time, before they eventually get released by a random fluctuation.
We call this the \textit{stuck/push} scenario.



\section{Results}

\subsection{How single cell motion depends on cell polarity}

\noindent 
Cell polarity is a fundamental feature that plays a key role in many cellular functions, such as cell growth, division, and migration.
In particular, depending on their degree of polarity, motile cells can display several different motion patterns, such as random, oscillatory or persistent movement.
For a motile cell, polarity is typically defined based on the leading and trailing edge (head and tail) of the moving cell.
In the model used here, it is the biochemical component $c_i$ which triggers the formation of membrane protrusions (pseudopods) at the cell boundary, and is thus responsible for setting the sense of orientation of the moving cell.
For this reason, we quantify polarity based on a polarity vector $\vec{P}$ that measures the asymmetry in the subcellular distribution of the biochemical component $c_i$, see Section~\ref{sec:polarity} for details.
To what extent the displacement of the cell is governed by the direction of the polarity vector is the subject of study of this section.

A wide range of cell motion patterns was studied in numerical simulations of our model previously~\cite{alonso2018modeling}.
It was shown that the transition from random to persistent motion is controlled by the model parameter $k_a$, see Eq.~(\ref{eqci}).
Small values of $k_a$ give rise to an erratic trajectory that is caused by the random appearance of protrusions all around the cell.
On the other hand, large values of $k_a$ produce a more persistent trajectory due to the accumulation of protrusions in one region of the membrane that sets the overall direction of motion.
Those scenarios were associated with the vegetative and the starvation-developed states of {\it D.~discoideum} cells~\cite{alonso2018modeling}, respectively. 

\begin{figure}[h!]
\begin{center}
\includegraphics[width=0.81\textwidth]{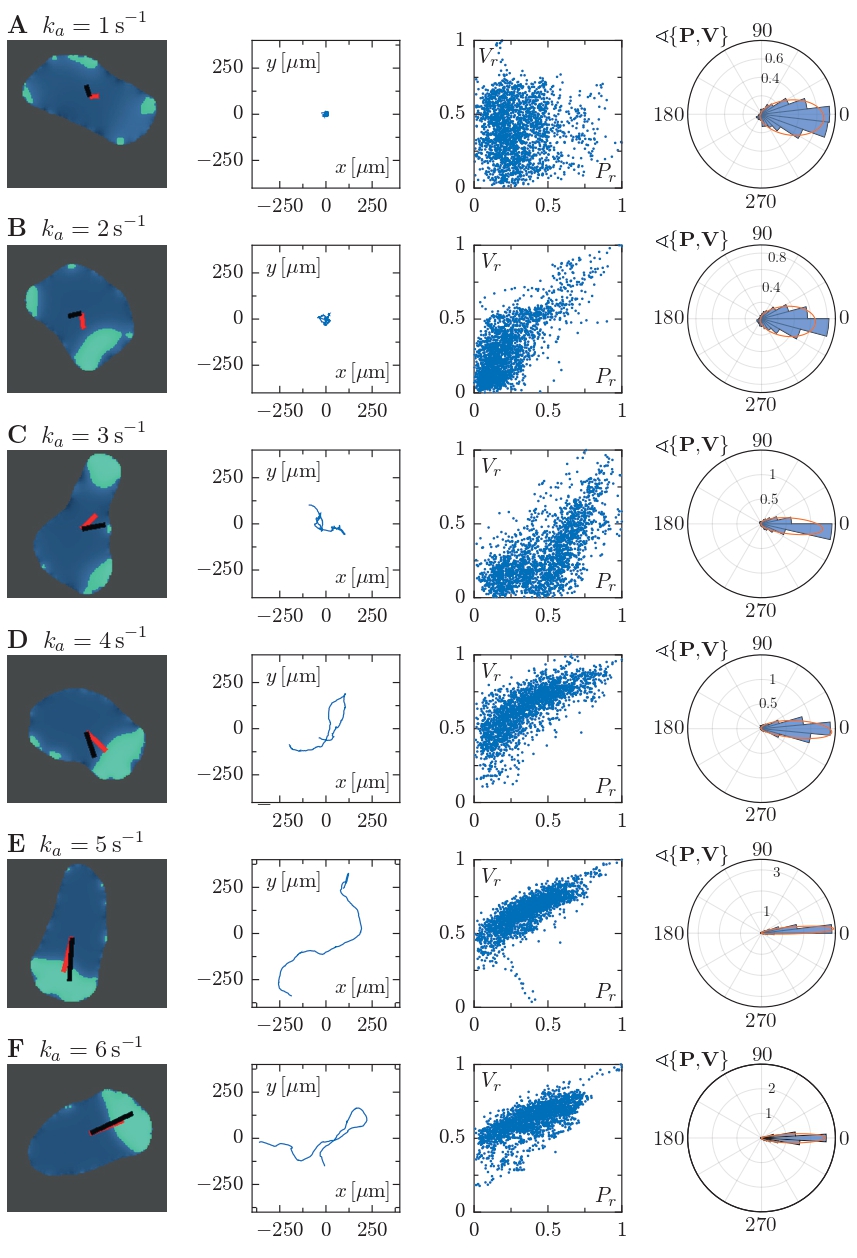}
\end{center}
\vspace{-0.4cm}
\caption{Each row corresponds to a particular single cell phenotype for increasing values of the parameter~$k_a$. In the first column, snapshots along with the polarization and velocity vectors $\vec{P}$ and $\vec{V}$ are depicted in black and red, respectively. The corresponding trajectory is shown in the second panels. Third panels represents a the correlation of the magnitudes of vectors $\vec{P}$ and~$\vec{V}$, which were normalized with respect to the maximal value observed during a simulation. The fourth column sub-panels represent circular histograms for the angle between the vectors~$\vec{P}$ and~$\vec{V}$.  Parameter values:~\textbf{A}~$k_a = 1 \, \mbox{s}^{-1}$, \textbf{B}~$k_a = 2 \, \mbox{s}^{-1}$, \textbf{C}~$k_a = 3 \, \mbox{s}^{-1}$, \textbf{D}~$k_a = 4 \, \mbox{s}^{-1}$, \textbf{E}~$k_a = 5 \, \mbox{s}^{-1}$, \textbf{F}~$k_a = 6 \, \mbox{s}^{-1}$. 
}\label{fig:2}
\end{figure}

Here, we studied cell migration patterns for different values of the parameter $k_a$, focusing on the dynamics of the displacement and polarity vectors to quantitatively analyze the role of cell polarization for the different modes of locomotion~(see Figure~\ref{fig:2}). 
For every row in Figure \ref{fig:2}, the panel in the first column shows a snapshot of a simulated cell with a superposition of the polarization vector $\vec{P}$ (in black) and the cell propagation vector $\vec{V}$ (in red).
The second column presents the cell trajectories in space.
Here, a pronounced change from random to persistent motion can be seen from top to bottom with growing $k_a$, in line with our earlier results~\cite{alonso2018modeling}.
Thus, cells perform larger explorations in space as they become more persistent for high values of the parameter $k_a$. 
In the third column, the correlation of the normalized magnitude of the vectors $\vec{P}$ and $\vec{V}$ is displayed, revealing a high degree of correlation for large values of~$k_a$, whereas correlations are low for small~$k_a$. 
In the fourth column, circular histograms of the angle $\theta$ between the vectors $\vec{P}$ and $\vec{V}$ are displayed.
For small values of $k_a$ both vectors are typically misaligned, while alignment is observed for high values of $k_a$.
The corresponding probability distribution exhibits a peak around zero for small values of the angle that becomes increasingly pronounced for growing values of $k_a$.
%
These results show that the vectors~$\vec{P}$ and~$\vec{V}$ are more likely to align for growing values of $k_a$, resulting in an increased probability for values of $\theta$ close to zero.

\begin{figure}[h!]
\begin{center}
\includegraphics[width=15cm]{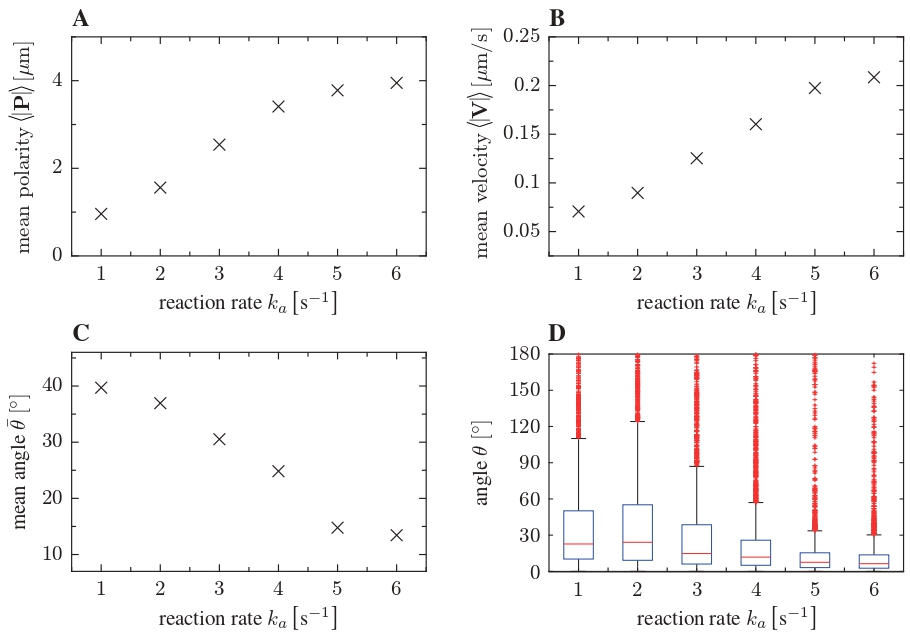}
\end{center}
\caption{\textbf{A} Magnitude of mean polarity $\vec{P}$, \textbf{B} mean velocity $\vec{V}$ and \textbf{C} the mean angle for different values of the reaction rate parameter~$k_a$. Accordingly, this parameter controls the polarity of single cells. \textbf{D} Box plots of the angle~$\theta$ as a function of the reaction rate parameter~$k_a$. For high~$k_a$, velocity and cell polarity tend to be aligned. }\label{fig:3}
\end{figure}

%
In short, the dynamics of the cell propagation vector~$\vec{V}$, the polarization vector~$\vec{P}$ and the angle~$\theta$ between them reflects the more persistent and less random motion of cells as the parameter value of~$k_a$ increases. 
To summarize the dependencies on the parameter $k_a$, we show the mean of the absolute values of the vectors $\vec{P}$ and $\vec{V}$ as well as the angle $\theta$ between them as a function of $k_a$ in Figure~\ref{fig:3}.
With increasing parameter~$k_a$, larger mean values of the polarization vector $\vec{P}$ are observed in Figure~\ref{fig:3}A.
This is associated with increasing mean values of the velocity of polarized cells, see the magnitudes of the vector $\vec{V}$ in Figure~\ref{fig:3}B.
Figure~\ref{fig:3}C finally shows that not only the absolute values of $\vec{P}$ and $\vec{V}$ increase, but also their alignment is more pronounced with increasing $k_a$, so that the mean angle $\theta$ decreases.
For a better comparison, box plots of the angle $\theta$ are presented in Figure~\ref{fig:3}D. 
In summary, persistent motion (high values of $k_a$) is characterized by larger magnitudes of the vectors $\vec{P}$ and $\vec{V}$ and by smaller angles $\theta$ between them, i.e. by increased alignment of the polarity and the displacement directions.
On the other hand, random motion (small values of $k_a$) results from smaller magnitudes of $\vec{P}$ and $\vec{V}$ and larger angles $\theta$, indicating the absence of alignment due to the irregular distribution of patches of $c_i$ that trigger random protrusion all around the cell boundary.


\subsection{Three scenarios of binary cell interactions}
\label{sec:3sc}

\begin{figure}[h!]
\begin{center}
\includegraphics[width=0.67\textwidth]{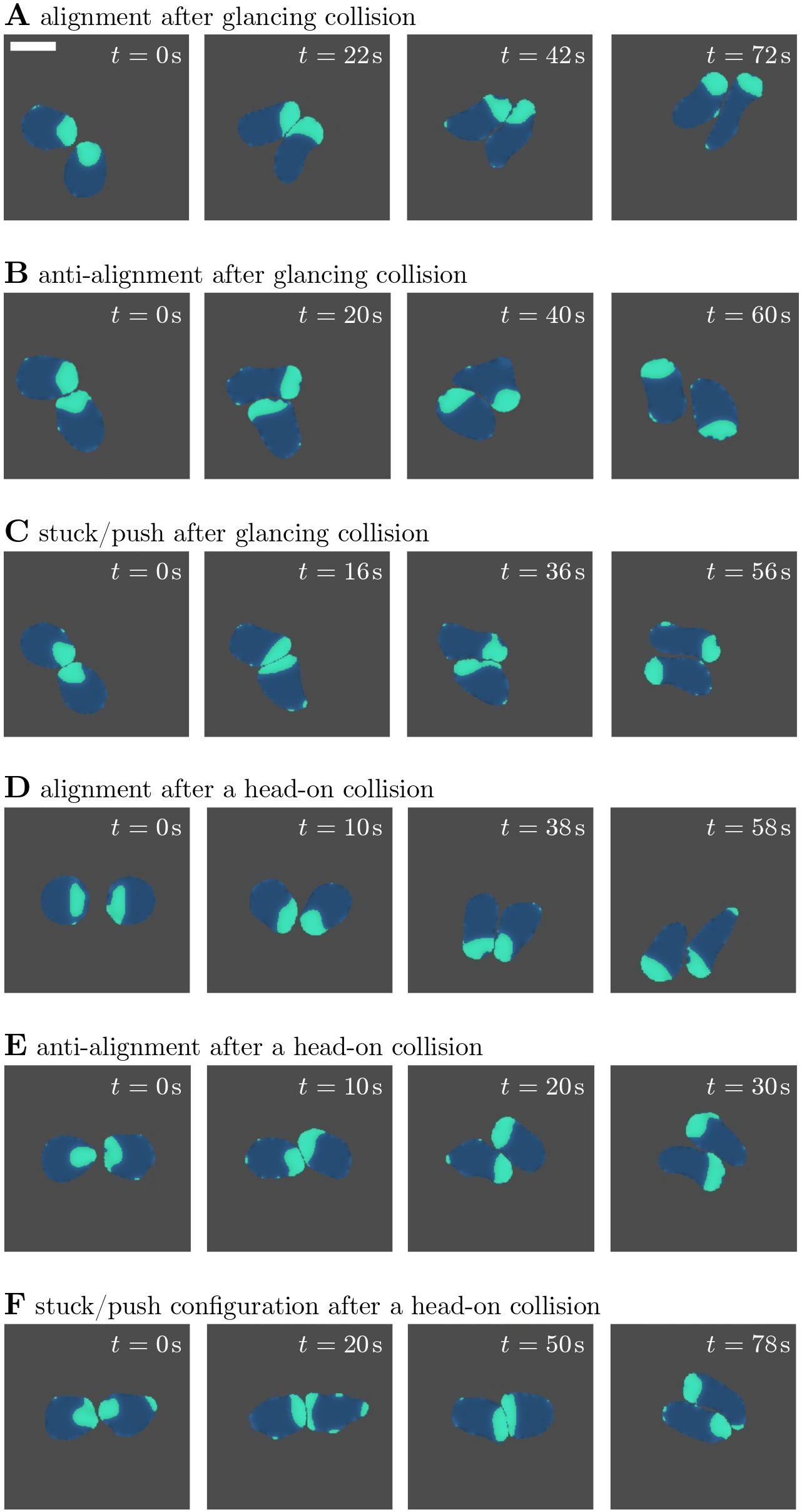}
\end{center}
\vspace{-0.3cm}
\caption{Sequence of snapshots of a binary interaction of cells, representing \textbf{A} alignment, \textbf{B} anti-alignment and \textbf{C} stuck/push scenarios for a glancing collision. Panels \textbf{D}-\textbf{F} show the alignment, anti-alignment and stuck/push scenarios, respectively, corresponding to head collision dynamics. Scale bar: $10\mu m$. }\label{fig:5}
\end{figure}

\noindent 
Before we address the dynamics of multiple interacting cells, we first focused on the interactions between two cells.
As detailed in the Section~\ref{sec:phasefield}, we extended the phase field description of a single amoeboid cell by including a repulsive force between them; any kind of adhesive force is neglected.
For the following simulations, we have chosen the model parameters in the regime of persistent motion, in particular, the reaction rate was set to $k_a=5s^{-1}$.
We modeled a pair of cells close to each other on a square grid for a period of $90\,\mbox{s}$ with the purpose of making them interact and analyzing the collision dynamics between them. 
From our simulations, we could distinguish three types of collision scenarios that we termed {\it alignment}, {\it anti-alignment} and {\it stuck/push}, illustrated in Figure~\ref{fig:4} (see also Section~\ref{sec:scenarios}).
In Figure~\ref{fig:5}, we display representative series of snapshots for each of the three scenarios observed in simulations of glancing collisions (Figure~\ref{fig:5}A-C) as well as for head-on collisions (Figure~\ref{fig:5}D-F).
We found that the avoidance between cells (anti-alignment) was the most frequently observed case in our simulations, see Figure~\ref{fig:5}B and~E.
The second most frequent scenario was the alignment of cells, as can be seen in Figure~\ref{fig:5}A and~D.
Finally, the stuck/push scenario was only rarely observed, see Figure~\ref{fig:5}C and~F. The two cells which get stuck upon collision, pushing against each other for an extended period of time, before the shape of one of them is strongly distorted, as a consequence of which the heads-on pushing configuration is destabilized and they continue moving in different directions.
However, as we will show in the next sections, this case will be more frequently observed when the density of cells is increased.

\begin{figure}[h!]
\begin{center}
\includegraphics[width=0.73\textwidth]{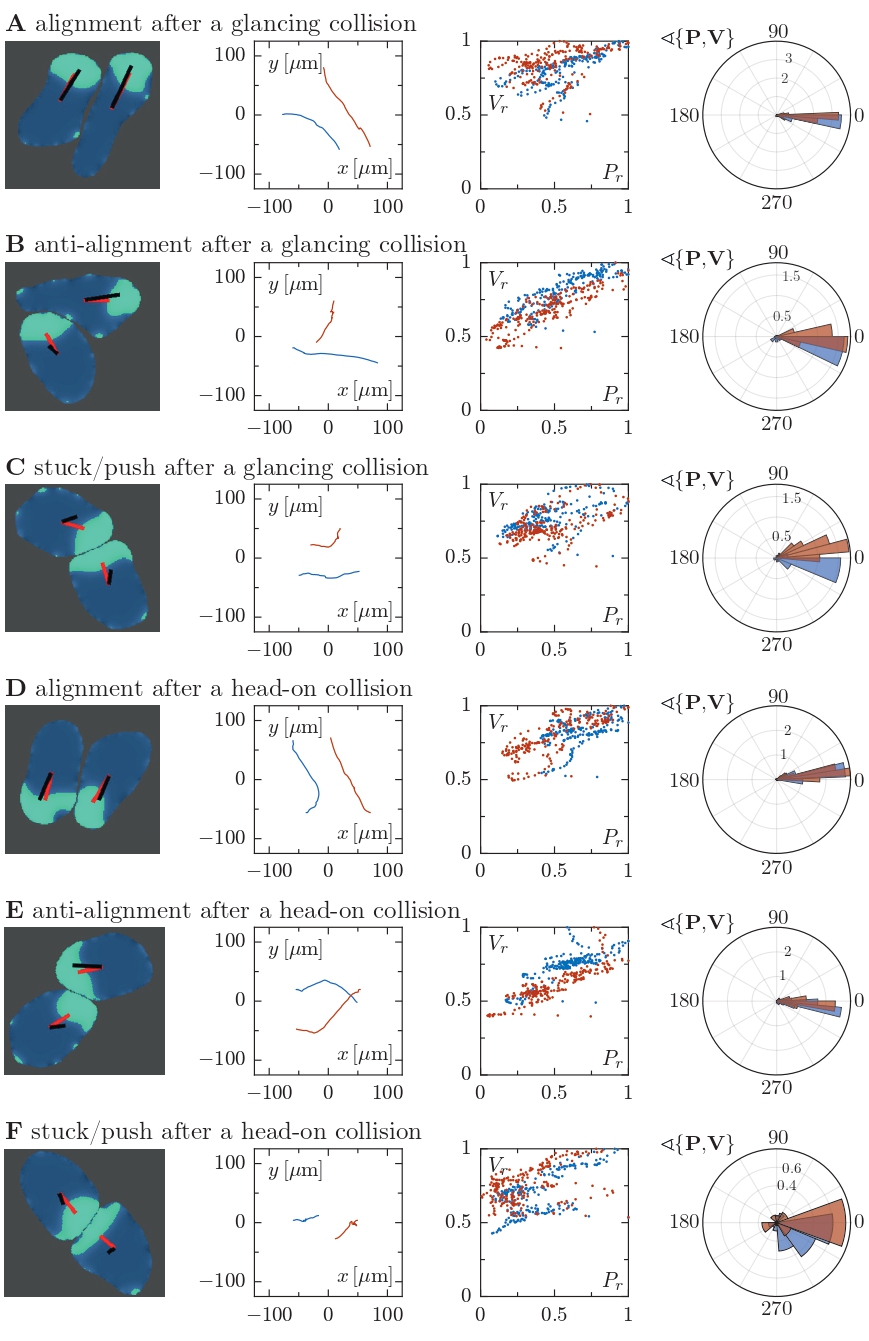}
\end{center}
\vspace{-0.5cm}
\caption{Quantitative analysis of binary cell interactions in terms of the cell polarity and velocity vectors for different interaction scenarios, cf.~Figure~\ref{fig:5}. For every row~\textbf{A}-\textbf{F}, the first column panels show the snapshots of cells where a representation of the vectors~$\vec{P}$ and~$\vec{V}$ is included. The second column shows the corresponding trajectories. Analogously to Figure~\ref{fig:2}, the third column represents the correlation of the rescaled magnitudes of vectors $\vec{P}$ and~$\vec{V}$ of the two interacting cells. The panels in fourth column show circular histograms of the angle~$\theta$ enclosed by cell polarity~$\vec{P}$ and velocity~$\vec{V}$. The collision scenarios in \textbf{A}-\textbf{C} correspond to glancing collisions, resulting in \textbf{A} alignment, \textbf{B} anti-alignment and \textbf{C} stuck/push configurations. In contrast, panels \textbf{D}-\textbf{F} correspond to head-on collisions, likewise resulting in \textbf{D} alignment, \textbf{E} anti-alignment and \textbf{F} stuck/push configurations.  }\label{fig:6}
\end{figure}

We furthermore analyzed the collision-induced dynamics of the cells in more detail in Figure~\ref{fig:6},
following the same order as in Figure~\ref{fig:5}.  
In the first column, we display snapshots of the two interacting cells for every studied case, including the polarity vector $\vec{P}$ and velocity vector $\vec{V}$ for each cell.
In the second column, representations of the cell trajectories are shown, revealing clear differences between the different collision scenarios.
In the alignment case (A and~D), for example, trajectories tend to be parallel, while a crossing of tracks is seen for the anti-alignment scenario (B and~E). 
Only small displacements are obtained in the stuck/push case (C and~F) as cells impede each others motion upon heads-on collision.
The third column displays again the correlation of the normalized magnitudes of the vectors $\vec{P}$ and $\vec{V}$. 
The fourth column shows the distributions of the angle $\theta$ between the polarity and velocity vectors $\vec{P}$ and $\vec{V}$ for the two interacting cells~--~wider distributions for the angle are seen in the stuck/push scenario.

\begin{figure}[h!]
\begin{center}
\includegraphics[width=0.85\textwidth]{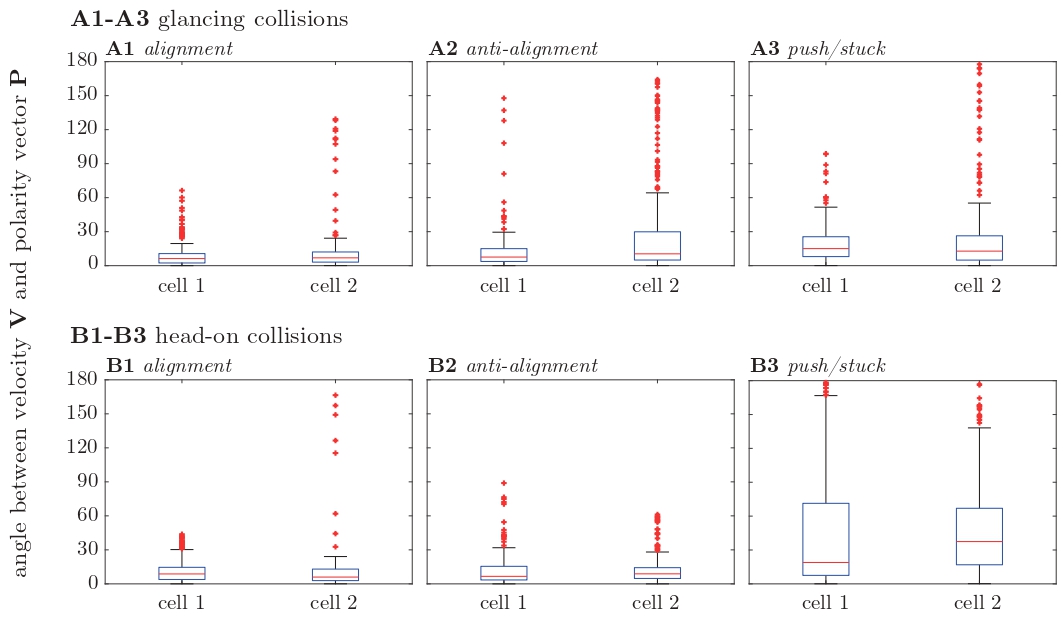}
\end{center}
\caption{Boxplot representations of the measured angles~$\theta$ during the interaction of two cells. The panels correspond to alignment, anti-alignment and stuck/push cases for the glancing collisions~(upper panels) and heads-on collisions~(lower panels), respectively. }\label{fig:7}
\end{figure}

For all the cases, the magnitude of both vectors tend to correlate, regardless of the orientation of the cells.
This behavior is more clearly observed in the alignment and anti-alignment cases, while in the stuck/push case this tendency is weaker. 
As mentioned before, these distributions are wider for the stuck/push case, indicating stronger fluctuations and less correlations in the orientations of the polarity and velocity vectors $\vec{P}$ and $\vec{V}$.
%
The results discussed above are summarized by box plots of the angle~$\theta$ between~$\vec{P}$ and~$\vec{V}$ for the different scenarios shown in Figure \ref{fig:7}. The alignment and anti-alignment scenarios generate only small differences in the orientations of the vectors $\vec{P}$ and $\vec{V}$ during the interactions and, therefore, $\vec{P}$ and $\vec{V}$ are typically aligned. In contrast, larger deviations of the relative orientation of the vectors and, consequently, larger values of $\theta$ occur in the stuck/push scenario, indicating that vectors $\vec{P}$ and $\vec{V}$ are not aligned.

We have furthermore used the cross correlation as a tool to show if a relationship between velocity and polarization vectors among the cells can be observed in order to clarify the results described above. However, the cross-correlation of the vectors $\vec{P}_1$ and $\vec{P}_2$, where the index indicates the different cells, has not revealed any conclusive information because of the different time scales of the repolarization, and the cross-correlation of vectors $\vec{V}_1$ and $\vec{V}_2$ only shows significant differences for the push conditions. The differences of collision types are most evident from the Supplementary Movies SM1-SM6 of the head-on and glancing binary collisions.












\noindent 
\subsection{Density dependence of effective cell motility and collective pattern formation}
\label{sec:dens_inc}

\noindent 

We now move to larger numbers of interacting cells. 
From the various cell-cell interactions within the ensemble of cells, we were able to identify the previously observed three collision scenarios: alignment, anti-alignment and stuck/push, see Supplementary Figure SF1.

\begin{figure}[h!]
\begin{center}
\includegraphics[width=\textwidth]{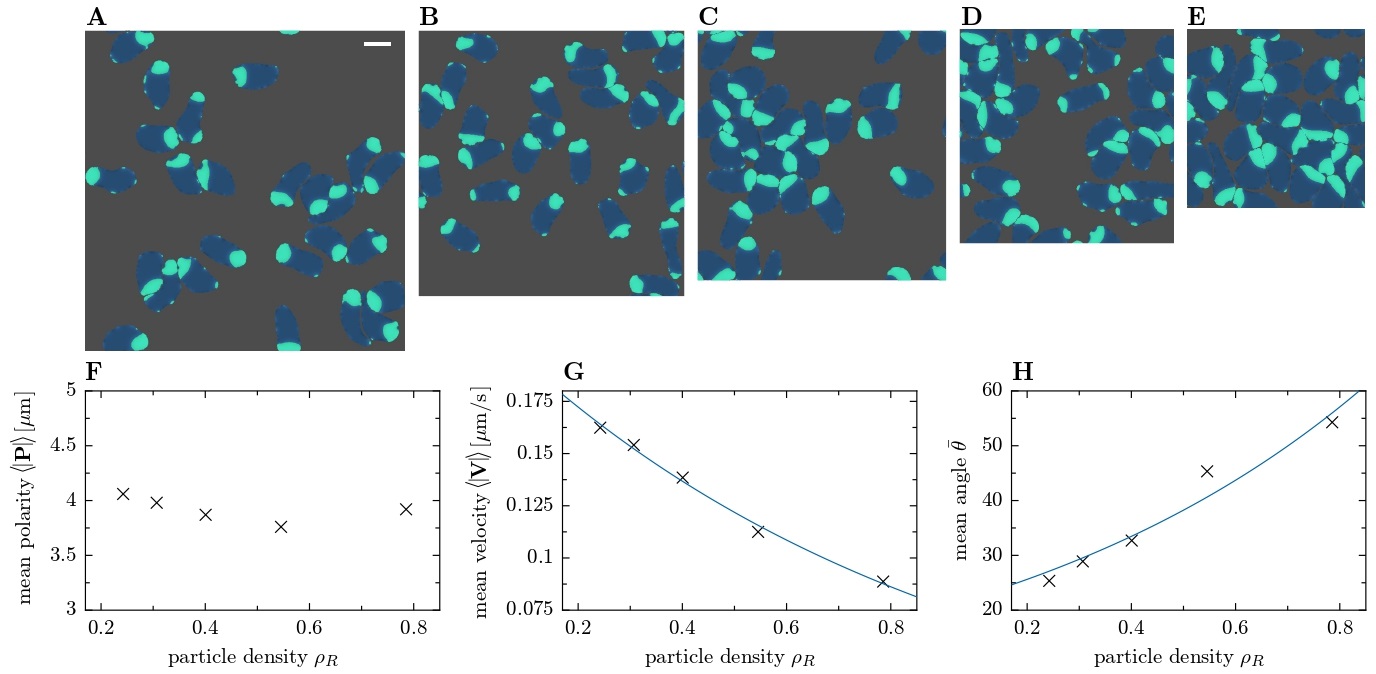}
\end{center}
\caption{Snapshots of simulations with constant cell number~$N=25$ and varying system size with a length separation of $L=12 \mu m$ between consecutive frames. The side length~$L$ of the simulation box and the particle density $\rho_R = NA_0/L^2$ are: \textbf{A}~$L=108 \mu m$ and $\rho_R=0.24$; \textbf{B}~$L=96 \mu m$ and $\rho_R=0.30$; \textbf{C}~$L=84 \mu m$ and $\rho_R=0.40$; \textbf{D}~$L=72 \mu m$ and $\rho_R=0.54$; \textbf{E}~$L=60 \mu m$ and $\rho_R=0.78$. \textbf{F} Mean polarity, \textbf{G} mean speed and \textbf{H} mean angle between velocity and polarity for different particle densities~$\rho_R$. The number of cells was fixed~($N=25$ cells), while the system size was varied. The fits in panels~\textbf{G} and~\textbf{H} provide a guide to the eye. }\label{fig:9}
\end{figure}

We varied the system size~$L$, keeping the number of cells~$N=25$ fixed, to assess the density dependence of motility characteristics, the relevance of collision scenarios and the resulting collective pattern formation. 
%
%
The cell density~$\rho_R = N A_0/L^2$ is determined by the number of cells per unit area in the simulation box. 
%
%
At high cell densities, jamming is expected to be more relevant as the probability of cell-cell collisions increases. 
%
%
To study these effects, we simulated five different grid sizes; Figure \ref{fig:9} shows exemplarily the gradual transition from a dilute~(Figure \ref{fig:9}A and Supplementary Movie SM7) to a dense system~(Figure \ref{fig:9}E and Supplementary Movie SM8).


In high density scenarios, stuck/push interactions are more commonly observed as cells trying to follow their own trajectory collide with others, thereby competing for free space. 
This behavior is expected due to the higher packing fraction. 
For the chosen model parameters, cells tend to remain fairly polarized~--~the mean magnitude of the polarity vector does only weakly depend on the cell density~(cf.~Figure \ref{fig:9}F). 
The mean velocity, however, decreases significantly as the cell density is increased~(see Figure \ref{fig:9}G) due to the lack of free space in a dense system. 
%
%
As a result of frequent cell-cell collisions, the angle between polarity and displacement vectors increases on average as a function of the density~$\rho_R$ as shown in Figure~\ref{fig:9}H.
This is in line with our previous observation that velocity~$\vec{V}$ and cell polarity vector~$\vec{P}$ tend to be misaligned as a result of stuck/push interactions~(see Figure~\ref{fig:6}F and the corresponding discussion in Section~\ref{sec:3sc}). 
%
%
%
As cells remain polarized~--~most of the biochemical component~$c_i$ is concentrated in one certain part of cells~--~but collisions change their velocities, the vectors $\vec{P}$ and $\vec{V}$ cease to be aligned and the angle between them, thus, increases. 

\begin{figure}[h!]
\begin{center}
\includegraphics[width=\textwidth]{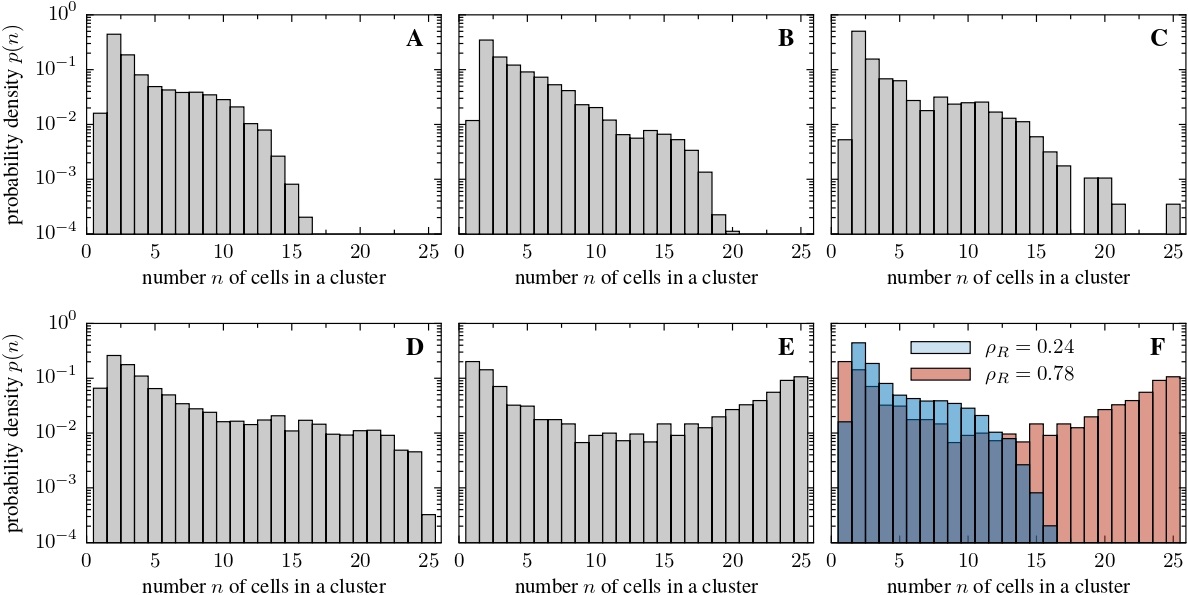}
\end{center}
\caption{Histograms of the observed cluster sizes for various particle densities. The cell number was kept constant~($N=25$ cells) and the system size was varied. The densities values of the different panels are \textbf{A}~$\rho_R=0.24$, \textbf{B}~$\rho_R=0.30$, \textbf{C}~$\rho_R=0.40$, \textbf{D}~$\rho_R=0.54$ and \textbf{E}~$\rho_R=0.78$. For increasing density, the cluster size distribution changes from a unimodal to a bimodal structure, signaling a clustering transition. This transition is highlighted in the last panel~\textbf{F}, where the cluster size distribution for low and high density is overlayed to simplify the comparison. }
\label{fig:10}
\end{figure}

The numerical results suggest that stuck/push collision scenarios are more relevant at higher cell densities. 
As cells impede each others motion as a result of stalling, these collisions may lead to the formation of clusters composed of immobile cells. 
This mechanism is reminiscent of the explanation of motility-induced phase separation~(MIPS) as observed in self-propelled discs~\cite{caes2015motility} that show phase separation at high density because particles hamper each others motion upon heads-on collision. 
We want to highlight, however, that the clustering dynamics observed in deformable cells is very different from classical MIPS:~as cells are deformable, the local stress acting on one cell is anisotropic.
This effect is even more pronounced as clusters grow in size and, consequently, the local pressure increases. 
As a result, clusters frequently break apart, thereby giving rise to a complex and dynamic clustering dynamics. 
To quantify clustering, we measured the cluster size distribution numerically~(Figure \ref{fig:10}). 
We consider two cells to belong to the same cluster if the distance between them is smaller than $12 \mu m$ which corresponds roughly to twice the average radius of a cell~(in the absence of interactions). 
At low densities, a few collisions lead to the transient formation of small groups of cells. 
Group sizes range from a few cells to more than half of the total number of cells in the system~(Figure \ref{fig:10}A).
In contrast, larger groups are more frequently formed in dense systems~(Figure \ref{fig:10}B-D). 
Accordingly, the cluster size distribution broadens as the cell density is increased and adopts even a bimodal shape in the high density limit~(Figure \ref{fig:10}E). 
A directed comparison of the transition from a unimodal to a bimodal shape can be seen in Figure \ref{fig:10}F, where panels corresponding to the lowest and highest density are shown together for a better visualization.
This structural change of the cluster size distribution eventually allows to define a critical density above which clustering sets in, similar to the clustering transition observed in ensembles of self-propelled rods~\cite{peruani2006nonequilibrium,bar2020self}.





\begin{figure}[h!]
\begin{center}
\includegraphics[width=\textwidth]{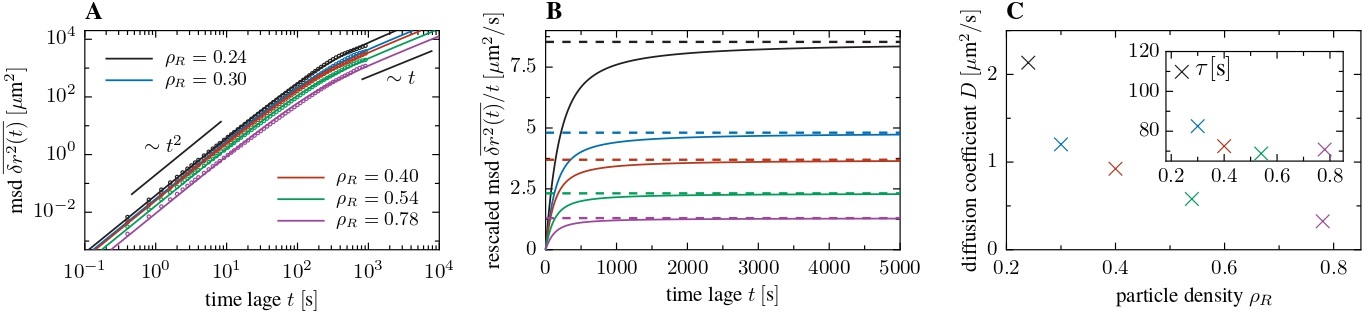}
\end{center}
\vspace{-0.2cm}
\caption{\textbf{A} Mean square displacement~(MSD) of particles for a system of~$N=25$ cells in a box of various side length~(implying various density). \textbf{B} Mean square displacement~(MSD) of particles, divided by time, for a system of~$N=25$ cells in a box of various side length. \textbf{C} Density dependence of the effective diffusion coefficient, extracted via fitting Fürths formula to the mean square displacement; the inset shows the reduction of the crossover timescale~$\tau$ as a function of the density~[Eq.~\eqref{eq:fuerth}]. }\label{fig:11}
\end{figure}

We furthermore quantify the random transport of cells within an ensemble by the mean square displacement~$ \langle \left | \Delta \vec{x}(t)\right |^2\rangle$, i.e.~the average displacement of a cell evaluated at different time lags. 
The mean square displacement~(MSD) decreases monotonically with the cell density~$\rho_R$ as cells can move more persistently at lower cell densities.
In call cases, we observe a transition from a ballistic regime~($\langle \left | \Delta \vec{x}(t)\right |^2\rangle \sim t^2$) at short time scales to a diffusive regime~($\langle \left | \Delta \vec{x}(t)\right |^2\rangle \sim t$) in the long-time limit ~(see Figure \ref{fig:11}). 
%
%
We measured the density dependence of the effective diffusion coefficient~$D$ of cells by fitting Fürths formula, 
%
\begin{linenomath}
\begin{align}
\label{eq:fuerth}
\left < \left | \vec{x}(t) - \vec{x}(t=0) \right |^2 \right > = 4D \! \left [ t - \tau \! \left ( 1 - e^{-t/\tau} \right ) \right ] \! ,
\end{align}
\end{linenomath}
to the numerically obtained MSD curves. 
The diffusion coefficient decreases as the density is increased~(cf.~Figure \ref{fig:11}C). 
Moreover, the fits show that the crossover timescale~$\tau$ decreases as the density of particle increases, due to a reduction of the mean-free path and a higher collision frequency at high density.




We double-checked that equivalent results are obtained when the density is increased by changing the particle number in a system of fixed size of~$L = 108 \mu m$~(data not shown) instead of fixing the particle number and decreasing the side length of the simulation box as discussed above~(Figures \ref{fig:9}-\ref{fig:11}). 

%
%



\subsection{Finite-size scaling reveals no significant system size dependence}

\begin{figure}[h!]
\begin{center}
\includegraphics[width=\textwidth]{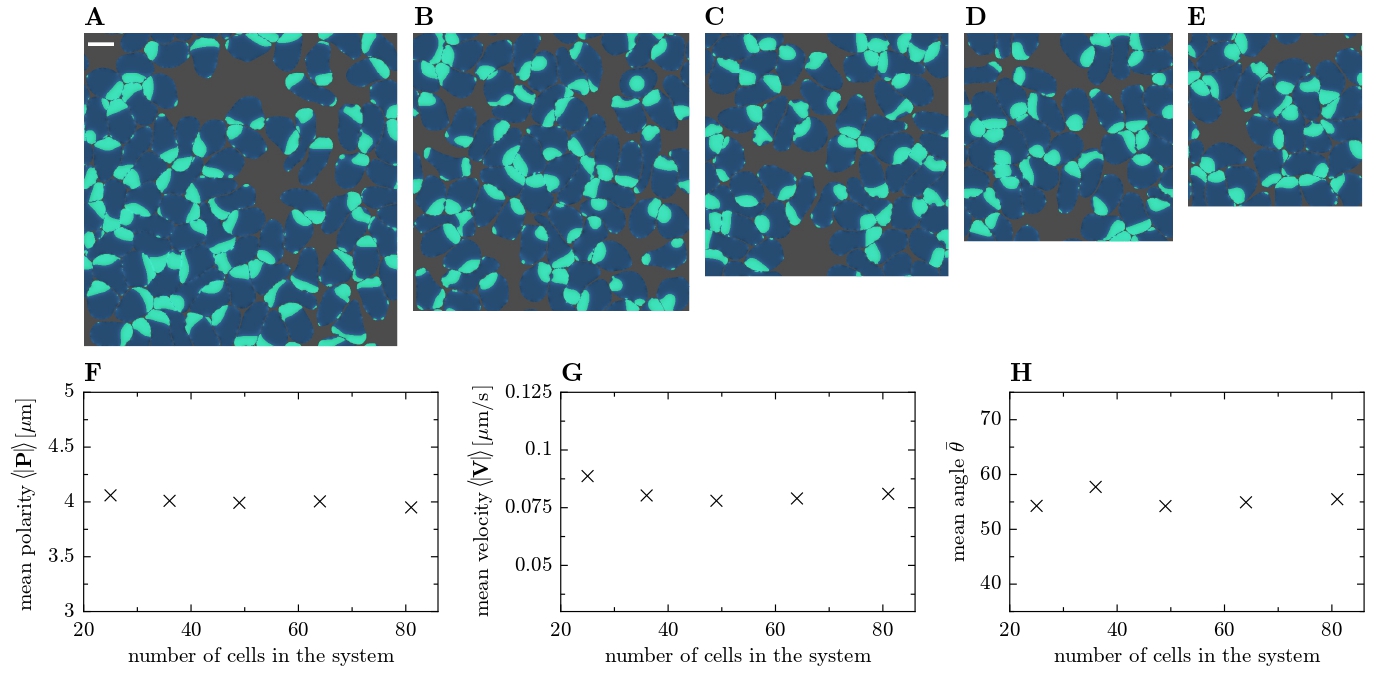}
\end{center}
\caption{Snapshots of simulations with the same particle density $\rho_R=0.78$. The number of cells and the grid size is \textbf{A} 81 cells and $L=108\mu m$, \textbf{B} 64 cells and $L=96\mu m$, \textbf{C} 49 cells and $L=84\mu m$, \textbf{D} 36 cells and $L=72\mu m$, \textbf{E} 25 cells and $L=60\mu m$. \textbf{F} Mean polarity, \textbf{G} mean velocity and \textbf{H} mean angle between velocity and polarity vector for a fixed particle density but various cell numbers in the system. }\label{fig:12}
\end{figure}

\noindent 
We conclude this section by assessing the relevance of the finite system size in numerical simulations. 
We performed a finite-size scaling analysis, varying the number of cells and system size such that the density or, equivalently, the packing fraction is kept constant. 
Five representative snapshots of the simulated systems are shown in Figure \ref{fig:12}A-E, where the number of cells is 25, 36, 49, 64 and 81, respectively, and the system size was adjusted from $L=60$ $\mu m$ to $L=108$ $\mu m$ accordingly. 
%
%
As shown in Figure \ref{fig:12}F and G, the measured value of the mean polarity and velocity does not reveal a significant system size dependence. 
Furthermore, the angle~$\theta$ between the vectors $\vec{P}$ and $\vec{V}$ in all cases does not change on average as the number of cells is varied~(cf.~Figure \ref{fig:12}H).

\begin{figure}[h!]
\begin{center}
\includegraphics[width=\textwidth]{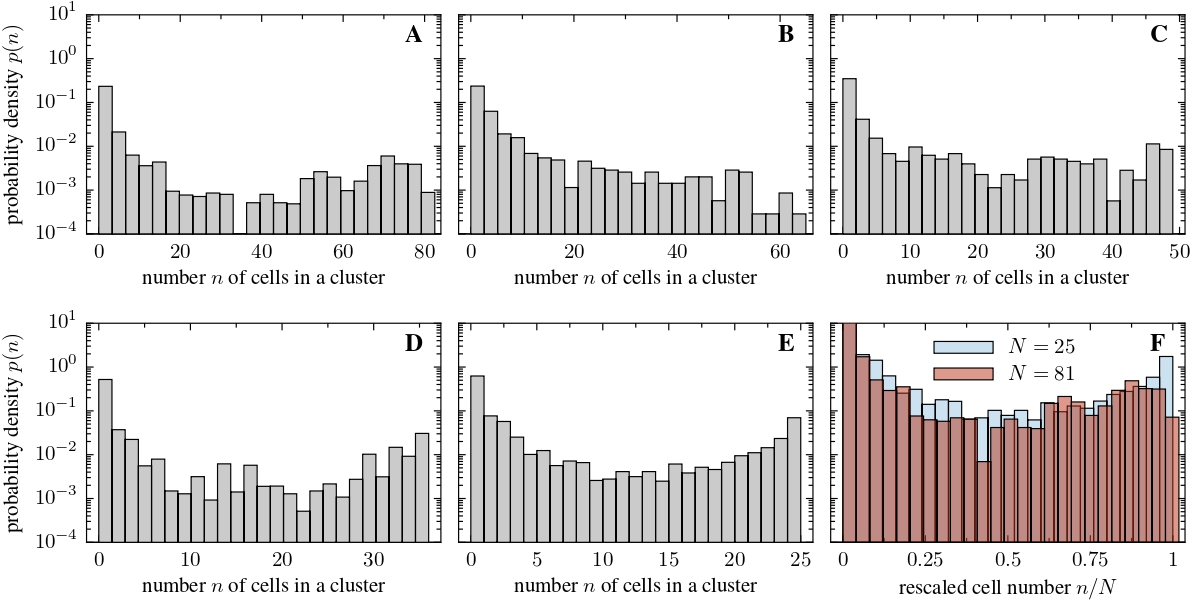}
\end{center}
\vspace{-0.4cm}
\caption{Cluster size distribution for a fixed density but various system sizes: \textbf{A} 81 cells and $L=108\mu m$, \textbf{B} 64 cells and $L=96\mu m$, \textbf{C} 49 cells and $L=84\mu m$, \textbf{D} 36 cells and $L=72\mu m$, \textbf{E} 25 cells and $L=60\mu m$. In panel \textbf{F} we overlay the probability distribution of the rescaled cluster size~$n/N$ for the smallest and largest particle number, indicating that the structure of the cluster size distribution is system size independent. }\label{fig:14}
\end{figure}

The tendency of cells to align or impede each others motion upon collision in stuck/push configurations implies a tendency towards cluster formation in the system (see Supplementary Movies SM9 and SM10 for simulations with 49 and 64 cells, respectively). 
As the system size increases, larger clusters of cells tend to form as shown in Figure~\ref{fig:12}. 
We quantitatively investigated the clustering dynamics by measuring the cluster size distribution in the stationary state; a comparison of all the studied cases is displayed in Figure \ref{fig:14}.
In contrast to the numerical experiment discussed in Section~\ref{sec:dens_inc}, where the density was increased, the shape of the cluster size distribution is now qualitatively independent of the system size as the cell number and system size is increased simultaneously such that the cell density remains constant. 
The cluster size distribution turns out to be bimodal for the considered density~(Figures~\ref{fig:14}A-E). Figure \ref{fig:14}F shows the probability of the rescaled cluster size for the system with the highest ($N=81$) and smallest number of cells~($N=25$). Furthermore, note that the results display large fluctuations and conclusions have to be drawn with caution.

\section{Discussion}


%
%

%
We have employed a generic mathematical model to describe the collective pattern formation of soft, deformable, self-driven cells. 
The model couples the intracellular biochemistry, responsible for cell polarization, the formation of membrane protrusions and, thus, for active motion, with a phase field which accounts for the current position and shape of the cell membrane. 
%
%
%
Each cell is described by an individual phase field. 
We include a repulsive interaction between cells into a previously established phase field model for individual cells~\cite{alonso2018modeling} in order to prevent them from overlapping. 
%
%
This framework enables us to address the complex interplay of dynamic particle shape, nonlinear repulsive interactions, self-propelled motion and the emergence of collective patterns in the context of active matter. 

Individual \textit{D.~discoideum} cells may aggregate following chemotactic signals to form  a multicellular structure \cite{bonner2015cellular}. At large spatial scales, the chemotactic concentration and the density of cells can be modelled using reaction-diffusion equations \cite{vasiev1994simulation}. In addition, there are several attempts to model the life cycle of \textit{D.~discoideum} from single cell shape changes and chemotaxis to collective behavior~\cite{de2017critical,palsson2000model}. 

%
Our modeling framework constitutes an active matter system composed of individual propellers, which are simple models of deformable cells like amoeba, e.g.~{\it D.~discoideum} cells. 
Whereas the role of particle shape~\cite{bar2020self,grossmann2020particle,peruani2006nonequilibrium} and the symmetries of interactions~\cite{vicsek1995novel,chate2020dry} have extensively been debated for active matter systems, the relevance of the deformability of particles has seldomly been addressed~\cite{menzel2012soft,lober2015collisions}. 
%
%
Oftentimes, active particles are considered as point-like objects or active spins~\cite{romanczuk2012active}, in the spirit of the seminal Vicsek model~\cite{vicsek1995novel,chate2020dry}, or as rigid objects such as active microswimmers~\cite{elgeti2015physics}, self-propelled discs~\cite{buttinoni2013dynamical} or rods~\cite{bar2020self}. 
%
%
The advantage of these approaches is their low computational cost, allowing to simulate several thousands of particles simultaneously. 
The computational cost associated with the incorporation of the deformability of particles, however, precludes the modeling of large amounts of individuals. 
We have arrived at around 100 cells in our simulations and, therefore, we are still far away from the limit considered in more theoretical studies. 
However, we made a first step towards connecting the dynamics at different scales, namely the intracellular pattern formation~--~responsible for the polarization of cells~--~and the collective dynamics of several individuals. 

As observed in the simulations, cell deformation is an important aspect included in this work. It permits to recreate scenarios closer to reality, for example in the study of motility in reduced spaces such as narrow vessels in the process of cancer metastasis \cite{keshavarz2021systematic,nagel_geometry-driven_2014}. This deformation is driven by the accumulation of a biochemical component inside the cell and by the interactions among the different cells present in the system. This characteristic is also observed when nonequilibrium stresses are built up during cluster formation and cells begin to deform. This is quite different to rigid spheres, where other mechanisms influence the motion.

%
The numerical simulations presented in this work reveal very rich collective pattern formation phenomena. 
For rigid active particles, the symmetry of individuals determines the symmetry of the interaction potential and, thereby, constrain the symmetries of emergent patterns: self-propelled discs may undergo dynamic clustering or motility-induced phase separation~\cite{buttinoni2013dynamical,theurkauff2012dynamic}; in the classical Vicsek model with polar alignment interaction, polarly ordered structures are observed at the macroscale~\cite{vicsek1995novel,chate2020dry}; elliptical self-propelled rods, in contrast, may form polar or nematic patterns, which may even dynamically coexist, depending on the strength of self-propulsion~\cite{huber2018emergence,grossmann2020particle}. Collective dynamics of cells is strongly determined by the interactions among them. While we consider here only repulsion, there are other phenomena involved in cell-to-cell communication \cite{rappel2016cell} and in particular cell-to-cell contact-inhibition \cite{zimmermann2016contact}. Such interactions can be easily included in our model. There are already some attempts to include deformation in collective cell dynamics. A protrusion region has been added to rigid spheres to simulate the deformation of the cells~\cite{coburn2013tactile} and some phase field models have been already implemented to describe oscillations in epithelial cells \cite{peyret2019sustained}.
Deformable cells, however, do not fall in any of these categories as the particle shape and, thus, the symmetry of the interactions, is a dynamic feature which is determined by intracellular biochemical processes as well as the complex, nonlinear interactions with other cells due to collisions that, in turn, induce additional cell shape changes. 
Accordingly, we found different interaction scenarios that may lead to alignment, anti-alignment and stuck/push configurations, whose relevance depends on the relative position and orientation of two cells before the collision event, the stiffness of particles as well as the global cell density.


The modeling framework decouples the actual displacement of a cell in space from the intracellular polarization dynamics. 
Therefore, we could establish cell polarity $\vec{P}$ and velocity $\vec{V}$ vectors based on the biochemical concentration patches and spatial displacements, respectively, enabling us to quantify the intricate, nonlinear coupling of cell polarization and motion. 
We first analyzed it at the single-cell level for different parameters sets which could, for example, represent vegetative or starvation conditions of~\textit{D.~discoideum} cells. 
Furthermore, we established how cell polarity and velocity vectors~--~as well as the angle between them~--~behave in ensembles of interacting cells and studied their density and system size dependence. 
%
%
For the considered model parameters, cell polarization depends only weakly on cell density, whereas the velocity of cells strongly decreases as the packing fraction is increased.

Our modeling framework allows to tune parameters such that the cell polarity is rather stable. 
The stability of the polarity axis, due to the intracellular pattern formation mechanisms described here, represents an effective memory on the polarity axis~\cite{van2021short}.
This memory precludes the formation of pseudopods at the side and at the rear of the cell~\cite{bosgraaf2009ordered} when cells follow a gradient of chemoattractant~\cite{andrew2007chemotaxis}. 
In this case, the polarity axis can be determined by the external gradient or by fitting an ellipsoid to the cell.  

In single cell systems, velocity and polarity of cells are directly correlated.
During persistent motion, most of the biochemical component inside the cell is accumulated behind the leading edge of the cell membrane; cell polarity and velocity vectors are aligned to a high degree, i.e.~the angle between them is small. 
A random reorientation takes places if various patches of the biochemical concentration exist inside the cell, thereby inducing a misalignment of cell polarity and velocity vector. 
%

In ensembles of interacting cells, a nontrivial coupling of cell polarization and velocity emerges dynamically due to cell-cell collisions. 
Even if parameters are tuned such that cells remain highly polarized~(starvation conditions of \textit{D.~discoideum}), there is a density dependent contribution to the interaction between cell polarization and velocity. 
%
%
In particular, jamming of particles misaligns polarity~$\vec{P}$ and velocity~$\vec{V}$; in the high density regime, stuck/push collisions during which cells stall each others motion are most prominent. 
Consequently, we observe transient clustering~--~clusters dynamically build-up and break apart.
The stability of clusters is crucially determined by the deformability of cells:~as cells are soft, the local anisotropic stress acting on one cell inside a cluster yields cell-shape changes which may eventually destabilize the entire cluster. 
%
%
%
In the low density regime, in contrast, cells can move freely and persistently, reflected by the tendency of cell polarization and velocity to align.


%
%
Several model parameters were fixed within this study. There is potentially a plethora of modes of motility to be explored by changing the level of noise intensity, for example. 
Moreover, the modeling framework can be extended in many nontrivial ways in the future, e.g.~including confinement, attractive interactions or collision-induced inhibition~\cite{kulawiak2016modeling,coburn2013tactile,lober2015collisions}. 
%
%
Specific modifications will enable to bring the modeling closer to a particular application, such as the study of collective cell motility in wound repair, immune response and tissue morphogenesis~\cite{deforet2014emergence,doxzen2013guidance}.

In summary, we have developed a generic modeling framework to computationally simulate the dynamics of active matter systems which are composed of deformable particles, such as amoeboid cells.
This method bridges the gap between the intracellular biochemical kinetics, in turn controlling the membrane activity of cells, the resulting locomotion of amoeboid cells and their emergent collective dynamics. 
Understanding self-organization processes of cells is of high relevance, for example for the aggregation processes of {\it D.~discoideum}, but also for collective cell dynamics during immune responses and cancer spreading. 
%
%




\section*{Conflict of Interest Statement}

The authors declare that the research was conducted in the absence of any commercial or financial relationships that could be construed as a potential conflict of interest.

\section*{Author Contributions}

E.M.~performed the numerical simulations; E.M., R.G., C.B.~and S.A.~contributed to the design and implementation of the research, to the analysis of the results and to the writing of the manuscript.


\section*{Funding}
S.A.~and E.M.~thank support from MICINN (Spain), and FEDER, Spain (European Union), under project PGC2018-095456-B-I00. E.M.~acknowledges also financial support from CONACYT.
C.B.~and R.G~acknowledge financial support by the Deutsche Forschungsgemeinschaft (DFG) -- Project-ID 318763901 -- SFB1294, project~B02.




\bibliographystyle{frontiersinHLTH&FPHY} 
\bibliography{test}

\begin{thebibliography}{79}
\expandafter\ifx\csname natexlab\endcsname\relax\def\natexlab#1{#1}\fi
\expandafter\ifx\csname urlstyle\endcsname\relax
  \expandafter\ifx\csname doi\endcsname\relax
  \def\doi#1{doi:\discretionary{}{}{}#1}\fi \else
  \expandafter\ifx\csname doi\endcsname\relax
  \def\doi{doi:\discretionary{}{}{}\begingroup \urlstyle{rm}\Url}\fi \fi
\expandafter\ifx\csname selectlanguage\endcsname\relax
  \def\selectlanguage#1{}\fi

\bibitem[{Vicsek and Zafeiris(2012)}]{vicsek2021collective}
Vicsek T, Zafeiris A.
\newblock Collective motion.
\newblock {\em Physics Reports\/} {\bf 517} (2012) 71--140.

\bibitem[{Friedl and Gilmour(2009)}]{friedl2009collective}
Friedl P, Gilmour D.
\newblock Collective cell migration in morphogenesis, regeneration and cancer.
\newblock {\em Nature Reviews Molecular Cell Biology\/} {\bf 10} (2009)
  445--457.

\bibitem[{Poujade et~al.(2007)Poujade, Grasland-Mongrain, Hertzog, Jouanneau,
  Chavrier, Ladoux et~al.}]{poujade2007collective}
Poujade M, Grasland-Mongrain E, Hertzog A, Jouanneau J, Chavrier P, Ladoux B,
  et~al.
\newblock Collective migration of an epithelial monolayer in response to a
  model wound.
\newblock {\em Proceedings of the National Academy of Sciences\/} {\bf 104}
  (2007) 15988--15993.

\bibitem[{Heisenberg and Bella{\"\i}che(2013)}]{heisenberg2013forces}
Heisenberg CP, Bella{\"\i}che Y.
\newblock Forces in tissue morphogenesis and patterning.
\newblock {\em Cell\/} {\bf 153} (2013) 948--962.

\bibitem[{Dormann and Weijer(2003)}]{dormann2003chemotactic}
Dormann D, Weijer CJ.
\newblock Chemotactic cell movement during development.
\newblock {\em Current Opinion in Genetics~\& Development\/} {\bf 13} (2003)
  358--364.

\bibitem[{Montell(2003)}]{montell2003border}
Montell DJ.
\newblock Border-cell migration: the race is on.
\newblock {\em Nature Reviews Molecular Cell Biology\/} {\bf 4} (2003) 13--24.

\bibitem[{Cheung and Ewald(2016)}]{cheung2016collective}
Cheung KJ, Ewald AJ.
\newblock A collective route to metastasis: seeding by tumor cell clusters.
\newblock {\em Science\/} {\bf 352} (2016) 167--169.

\bibitem[{Friedl et~al.(2012)Friedl, Locker, Sahai, and
  Segall}]{friedl2012classifying}
Friedl P, Locker J, Sahai E, Segall JE.
\newblock Classifying collective cancer cell invasion.
\newblock {\em Nature Cell Biology\/} {\bf 14} (2012) 777--783.

\bibitem[{Blanchoin et~al.(2014)Blanchoin, Boujemaa-Paterski, Sykes, and
  Plastino}]{blanchoin2014actin}
Blanchoin L, Boujemaa-Paterski R, Sykes C, Plastino J.
\newblock Actin dynamics, architecture, and mechanics in cell motility.
\newblock {\em Physiological Reviews\/} {\bf 94} (2014) 235--263.

\bibitem[{Mitchison and Cramer(1996)}]{mitchison1996actin}
Mitchison T, Cramer L.
\newblock Actin-based cell motility and cell locomotion.
\newblock {\em Cell\/} {\bf 84} (1996) 371--379.

\bibitem[{Alberts et~al.(2015)Alberts, Johnson, Lewis, Morgan, Raff, Roberts
  et~al.}]{alberts1989molecular}
Alberts B, Johnson A, Lewis J, Morgan D, Raff M, Roberts K, et~al.
\newblock {\em Molecular {B}iology of the {C}ell\/} (W.~W.~Norton~\& Company)
  (2015).

\bibitem[{Farhadifar et~al.(2007)Farhadifar, R{\"o}per, Aigouy, Eaton, and
  J{\"u}licher}]{farhadifar2007influence}
Farhadifar R, R{\"o}per JC, Aigouy B, Eaton S, J{\"u}licher F.
\newblock The influence of cell mechanics, cell-cell interactions, and
  proliferation on epithelial packing.
\newblock {\em Current Biology\/} {\bf 17} (2007) 2095--2104.

\bibitem[{Kabla(2012)}]{kabla2012collective}
Kabla AJ.
\newblock Collective cell migration: leadership, invasion and segregation.
\newblock {\em Journal of the Royal Society Interface\/} {\bf 9} (2012)
  3268--3278.

\bibitem[{Petitjean et~al.(2010)Petitjean, Reffay, Grasland-Mongrain, Poujade,
  Ladoux, Buguin et~al.}]{petitjean2010velocity}
Petitjean L, Reffay M, Grasland-Mongrain E, Poujade M, Ladoux B, Buguin A,
  et~al.
\newblock Velocity fields in a collectively migrating epithelium.
\newblock {\em Biophysical Journal\/} {\bf 98} (2010) 1790--1800.

\bibitem[{Aranson(2016)}]{aranson2016physical}
Aranson IS, editor.
\newblock {\em Physical Models of Cell Motility\/} (Springer) (2016).

\bibitem[{Alert and Trepat(2020)}]{alert2020physical}
Alert R, Trepat X.
\newblock Physical models of collective cell migration.
\newblock {\em Annual Review of Condensed Matter Physics\/} {\bf 11} (2020)
  77--101.

\bibitem[{Camley and Rappel(2017)}]{camley2017physical}
Camley BA, Rappel WJ.
\newblock Physical models of collective cell motility:~from cell to tissue.
\newblock {\em Journal of Physics D:~Applied Physics\/} {\bf 50} (2017) 113002.

\bibitem[{Vicsek et~al.(1995)Vicsek, Czir{\'o}k, Ben-Jacob, Cohen, and
  Shochet}]{vicsek1995novel}
Vicsek T, Czir{\'o}k A, Ben-Jacob E, Cohen I, Shochet O.
\newblock Novel type of phase transition in a system of self-driven particles.
\newblock {\em Physical Review Letters\/} {\bf 75} (1995) 1226.

\bibitem[{Gr\'egoire and Chat\'e(2004)}]{gregoire2004onset}
Gr\'egoire G, Chat\'e H.
\newblock Onset of collective and cohesive motion.
\newblock {\em Physical Review Letters\/} {\bf 92} (2004) 025702.

\bibitem[{B{\"a}r et~al.(2020)B{\"a}r, Gro{\ss}mann, Heidenreich, and
  Peruani}]{bar2020self}
B{\"a}r M, Gro{\ss}mann R, Heidenreich S, Peruani F.
\newblock Self-propelled rods:~insights and perspectives for active matter.
\newblock {\em Annual Review of Condensed Matter Physics\/} {\bf 11} (2020)
  441--466.

\bibitem[{Gro{\ss}mann et~al.(2020)Gro{\ss}mann, Aranson, and
  Peruani}]{grossmann2020particle}
Gro{\ss}mann R, Aranson IS, Peruani F.
\newblock A particle-field approach bridges phase separation and collective
  motion in active matter.
\newblock {\em Nature Communications\/} {\bf 11} (2020) 5365.

\bibitem[{Buttinoni et~al.(2013)Buttinoni, Bialk\'e, K\"ummel, L\"owen,
  Bechinger, and Speck}]{buttinoni2013dynamical}
Buttinoni I, Bialk\'e J, K\"ummel F, L\"owen H, Bechinger C, Speck T.
\newblock Dynamical clustering and phase separation in suspensions of
  self-propelled colloidal particles.
\newblock {\em Physical Review Letters\/} {\bf 110} (2013) 238301.

\bibitem[{Cates and Tailleur(2015)}]{caes2015motility}
Cates ME, Tailleur J.
\newblock Motility-induced phase separation.
\newblock {\em Annual Review of Condensed Matter Physics\/} {\bf 6} (2015)
  219--244.

\bibitem[{Menzel and Ohta(2012)}]{menzel2012soft}
Menzel AM, Ohta T.
\newblock Soft deformable self-propelled particles.
\newblock {\em Europhysics Letters\/} {\bf 99} (2012) 58001.

\bibitem[{Glazier and Graner(1993)}]{glazier1993simulation}
Glazier JA, Graner F.
\newblock Simulation of the differential adhesion driven rearrangement of
  biological cells.
\newblock {\em Physical Review E\/} {\bf 47} (1993) 2128.

\bibitem[{Graner and Glazier(1992)}]{graner1992simulation}
Graner F, Glazier JA.
\newblock Simulation of biological cell sorting using a two-dimensional
  extended {P}otts model.
\newblock {\em Physical Review Letters\/} {\bf 69} (1992) 2013.

\bibitem[{Nagai and Honda(2001)}]{nagai2001dynamic}
Nagai T, Honda H.
\newblock A dynamic cell model for the formation of epithelial tissues.
\newblock {\em Philosophical Magazine B\/} {\bf 81} (2001) 699--719.

\bibitem[{Staple et~al.(2010)Staple, Farhadifar, R{\"o}per, Aigouy, Eaton, and
  J{\"u}licher}]{staple2010mechanics}
Staple DB, Farhadifar R, R{\"o}per JC, Aigouy B, Eaton S, J{\"u}licher F.
\newblock Mechanics and remodelling of cell packings in epithelia.
\newblock {\em The European Physical Journal E\/} {\bf 33} (2010) 117--127.

\bibitem[{Li and Sun(2014)}]{li2014coherent}
Li B, Sun SX.
\newblock Coherent motions in confluent cell monolayer sheets.
\newblock {\em Biophysical Journal\/} {\bf 107} (2014) 1532--1541.

\bibitem[{Bi et~al.(2016)Bi, Yang, Marchetti, and Manning}]{bi2016motility}
Bi D, Yang X, Marchetti MC, Manning ML.
\newblock Motility-driven glass and jamming transitions in biological tissues.
\newblock {\em Physical Review X\/} {\bf 6} (2016) 021011.

\bibitem[{Alt et~al.(2017)Alt, Ganguly, and Salbreux}]{alt2017vertex}
Alt S, Ganguly P, Salbreux G.
\newblock Vertex models: from cell mechanics to tissue morphogenesis.
\newblock {\em Philosophical Transactions of the Royal Society B: Biological
  Sciences\/} {\bf 372} (2017) 20150520.

\bibitem[{Kockelkoren et~al.(2003)Kockelkoren, Levine, and
  Rappel}]{kockelkoren2003computational}
Kockelkoren J, Levine H, Rappel WJ.
\newblock Computational approach for modeling intra-and extracellular dynamics.
\newblock {\em Physical Review E\/} {\bf 68} (2003) 037702.

\bibitem[{Camley et~al.(2013)Camley, Zhao, Li, Levine, and
  Rappel}]{camley2013periodic}
Camley BA, Zhao Y, Li B, Levine H, Rappel WJ.
\newblock Periodic migration in a physical model of cells on micropatterns.
\newblock {\em Physical Review Letters\/} {\bf 111} (2013) 158102.

\bibitem[{Shao et~al.(2010)Shao, Rappel, and Levine}]{shao2010computational}
Shao D, Rappel WJ, Levine H.
\newblock Computational model for cell morphodynamics.
\newblock {\em Physical Review Letters\/} {\bf 105} (2010) 108104.

\bibitem[{Ziebert et~al.(2012)Ziebert, Swaminathan, and
  Aranson}]{ziebert2012model}
Ziebert F, Swaminathan S, Aranson IS.
\newblock Model for self-polarization and motility of keratocyte fragments.
\newblock {\em Journal of the Royal Society Interface\/} {\bf 9} (2012)
  1084--1092.

\bibitem[{Shao et~al.(2012)Shao, Levine, and Rappel}]{shao2012coupling}
Shao D, Levine H, Rappel WJ.
\newblock Coupling actin flow, adhesion, and morphology in a computational cell
  motility model.
\newblock {\em Proceedings of the National Academy of Sciences\/} {\bf 109}
  (2012) 6851--6856.

\bibitem[{L{\"o}ber et~al.(2014)L{\"o}ber, Ziebert, and
  Aranson}]{lober2014modeling}
L{\"o}ber J, Ziebert F, Aranson IS.
\newblock Modeling crawling cell movement on soft engineered substrates.
\newblock {\em Soft Matter\/} {\bf 10} (2014) 1365--1373.

\bibitem[{Camley et~al.(2017)Camley, Zhao, Li, Levine, and
  Rappel}]{camley2017crawling}
Camley BA, Zhao Y, Li B, Levine H, Rappel WJ.
\newblock Crawling and turning in a minimal reaction-diffusion cell motility
  model:~coupling cell shape and biochemistry.
\newblock {\em Physical Review E\/} {\bf 95} (2017) 012401.

\bibitem[{Najem and Grant(2013)}]{najem2013phase}
Najem S, Grant M.
\newblock Phase-field approach to chemotactic driving of neutrophil
  morphodynamics.
\newblock {\em Physical Review E\/} {\bf 88} (2013) 034702.

\bibitem[{Alonso et~al.(2018)Alonso, Stange, and Beta}]{alonso2018modeling}
Alonso S, Stange M, Beta C.
\newblock Modeling random crawling, membrane deformation and intracellular
  polarity of motile amoeboid cells.
\newblock {\em PLoS ONE\/} {\bf 13} (2018) e0201977.

\bibitem[{Moure and Gomez(2016)}]{moure2016computational}
Moure A, Gomez H.
\newblock Computational model for amoeboid motion:~coupling membrane and
  cytosol dynamics.
\newblock {\em Physical Review E\/} {\bf 94} (2016) 042423.

\bibitem[{Moure and Gomez(2018)}]{moure2018three}
Moure A, Gomez H.
\newblock Three-dimensional simulation of obstacle-mediated chemotaxis.
\newblock {\em Biomechanics and Modeling in Mechanobiology\/} {\bf 17} (2018)
  1243--1268.

\bibitem[{Taniguchi et~al.(2013)Taniguchi, Ishihara, Oonuki, Honda-Kitahara,
  Kaneko, and Sawai}]{taniguchi2013phase}
Taniguchi D, Ishihara S, Oonuki T, Honda-Kitahara M, Kaneko K, Sawai S.
\newblock Phase geometries of two-dimensional excitable waves govern
  self-organized morphodynamics of amoeboid cells.
\newblock {\em Proceedings of the National Academy of Sciences\/} {\bf 110}
  (2013) 5016--5021.

\bibitem[{Flemming et~al.(2020)Flemming, Font, Alonso, and
  Beta}]{flemming2020cortical}
Flemming S, Font F, Alonso S, Beta C.
\newblock How cortical waves drive fission of motile cells.
\newblock {\em Proceedings of the National Academy of Sciences\/} {\bf 117}
  (2020) 6330--6338.

\bibitem[{Coburn et~al.(2013)Coburn, Cerone, Torney, Couzin, and
  Neufeld}]{coburn2013tactile}
Coburn L, Cerone L, Torney C, Couzin ID, Neufeld Z.
\newblock Tactile interactions lead to coherent motion and enhanced chemotaxis
  of migrating cells.
\newblock {\em Physical Biology\/} {\bf 10} (2013) 046002.

\bibitem[{Nonomura(2012)}]{nonomura2012study}
Nonomura M.
\newblock Study on multicellular systems using a phase field model.
\newblock {\em PLoS ONE\/} {\bf 7} (2012) e33501.

\bibitem[{Palmieri et~al.(2015)Palmieri, Bresler, Wirtz, and
  Grant}]{palmieri2015multiple}
Palmieri B, Bresler Y, Wirtz D, Grant M.
\newblock Multiple scale model for cell migration in monolayers:~elastic
  mismatch between cells enhances motility.
\newblock {\em Scientific Reports\/} {\bf 5} (2015) 11745.

\bibitem[{Peyret et~al.(2019)Peyret, Mueller, d’Alessandro, Begnaud, Marcq,
  M{\`e}ge et~al.}]{peyret2019sustained}
Peyret G, Mueller R, d’Alessandro J, Begnaud S, Marcq P, M{\`e}ge RM, et~al.
\newblock Sustained oscillations of epithelial cell sheets.
\newblock {\em Biophysical Journal\/} {\bf 117} (2019) 464--478.

\bibitem[{Mueller et~al.(2019)Mueller, Yeomans, and
  Doostmohammadi}]{mueller2019emergence}
Mueller R, Yeomans JM, Doostmohammadi A.
\newblock Emergence of active nematic behavior in monolayers of isotropic
  cells.
\newblock {\em Physical Review Letters\/} {\bf 122} (2019) 048004.

\bibitem[{Zhang et~al.(2020)Zhang, Mueller, Doostmohammadi, and
  Yeomans}]{zhang2020active}
Zhang G, Mueller R, Doostmohammadi A, Yeomans JM.
\newblock Active inter-cellular forces in collective cell motility.
\newblock {\em Journal of the Royal Society Interface\/} {\bf 17} (2020)
  20200312.

\bibitem[{Loewe et~al.(2020)Loewe, Chiang, Marenduzzo, and
  Marchetti}]{loewe2020solid}
Loewe B, Chiang M, Marenduzzo D, Marchetti MC.
\newblock Solid-liquid transition of deformable and overlapping active
  particles.
\newblock {\em Physical Review Letters\/} {\bf 125} (2020) 038003.

\bibitem[{Camley et~al.(2014)Camley, Zhang, Zhao, Li, Ben-Jacob, Levine
  et~al.}]{camley2014polarity}
Camley BA, Zhang Y, Zhao Y, Li B, Ben-Jacob E, Levine H, et~al.
\newblock Polarity mechanisms such as contact inhibition of locomotion regulate
  persistent rotational motion of mammalian cells on micropatterns.
\newblock {\em Proceedings of the National Academy of Sciences\/} {\bf 111}
  (2014) 14770--14775.

\bibitem[{L{\"o}ber et~al.(2015)L{\"o}ber, Ziebert, and
  Aranson}]{lober2015collisions}
L{\"o}ber J, Ziebert F, Aranson IS.
\newblock Collisions of deformable cells lead to collective migration.
\newblock {\em Scientific Reports\/} {\bf 5} (2015) 9172.

\bibitem[{Kulawiak et~al.(2016)Kulawiak, Camley, and
  Rappel}]{kulawiak2016modeling}
Kulawiak DA, Camley BA, Rappel WJ.
\newblock Modeling contact inhibition of locomotion of colliding cells
  migrating on micropatterned substrates.
\newblock {\em PLoS Computational Biology\/} {\bf 12} (2016) e1005239.

\bibitem[{Gregor et~al.(2010)Gregor, Fujimoto, Masaki, and
  Sawai}]{gregor2010onset}
Gregor T, Fujimoto K, Masaki N, Sawai S.
\newblock The onset of collective behavior in social amoebae.
\newblock {\em Science\/} {\bf 328} (2010) 1021--1025.

\bibitem[{Siegert and Weijer(1989)}]{siegert1989digital}
Siegert F, Weijer C.
\newblock {Digital image processing of optical density wave propagation in
  \textit{Dictyostelium discoideum} and analysis of the effects of caffeine and
  ammonia}.
\newblock {\em Journal of Cell Science\/} {\bf 93} (1989) 325--335.

\bibitem[{Grace and H{\"u}tt(2015)}]{grace2015regulation}
Grace M, H{\"u}tt MT.
\newblock {Regulation of spatiotemporal patterns by biological variability:
  General principles and applications to \textit{Dictyostelium discoideum}}.
\newblock {\em PLoS Computational Biology\/} {\bf 11} (2015) e1004367.

\bibitem[{Vidal-Henriquez and Gholami(2019)}]{vidal2019spontaneous}
Vidal-Henriquez E, Gholami A.
\newblock {Spontaneous center formation in \textit{Dictyostelium discoideum}}.
\newblock {\em Scientific Reports\/} {\bf 9} (2019) 3935.

\bibitem[{Moreno et~al.(2020)Moreno, Flemming, Font, Holschneider, Beta, and
  Alonso}]{moreno2020modeling}
Moreno E, Flemming S, Font F, Holschneider M, Beta C, Alonso S.
\newblock Modeling cell crawling strategies with a bistable model:~from
  amoeboid to fan-shaped cell motion.
\newblock {\em Physica D:~Nonlinear Phenomena\/}  (2020) 132591.

\bibitem[{Park et~al.(2017)Park, Holmes, Lee, Kim, Kim, Kwak
  et~al.}]{park2016mechano}
Park J, Holmes WR, Lee SH, Kim HN, Kim DH, Kwak MK, et~al.
\newblock Mechanochemical feedback underlies coexistence of qualitatively
  distinct cell polarity patterns within diverse cell populations.
\newblock {\em Proceedings of the National Academy of Sciences\/} {\bf 114}
  (2017) E5750--E5759.

\bibitem[{Peruani et~al.(2006)Peruani, Deutsch, and
  B\"ar}]{peruani2006nonequilibrium}
Peruani F, Deutsch A, B\"ar M.
\newblock Nonequilibrium clustering of self-propelled rods.
\newblock {\em Physical Review E\/} {\bf 74} (2006) 030904.

\bibitem[{Bonner(2015)}]{bonner2015cellular}
Bonner JT.
\newblock {\em Cellular slime molds\/} (Princeton University Press) (2015).

\bibitem[{Vasiev et~al.(1994)Vasiev, Hogeweg, and
  Panfilov}]{vasiev1994simulation}
Vasiev B, Hogeweg P, Panfilov A.
\newblock Simulation of dictyostelium discoideum aggregation via
  reaction-diffusion model.
\newblock {\em Physical Review Letters\/} {\bf 73} (1994) 3173.

\bibitem[{De~Palo et~al.(2017)De~Palo, Yi, and Endres}]{de2017critical}
De~Palo G, Yi D, Endres RG.
\newblock {A critical-like collective state leads to long-range cell
  communication in \textit{Dictyostelium discoideum} aggregation}.
\newblock {\em PLoS Biology\/} {\bf 15} (2017) e1002602.

\bibitem[{Palsson and Othmer(2000)}]{palsson2000model}
Palsson E, Othmer HG.
\newblock A model for individual and collective cell movement in dictyostelium
  discoideum.
\newblock {\em Proceedings of the National Academy of Sciences\/} {\bf 97}
  (2000) 10448--10453.

\bibitem[{Chat\'{e}(2020)}]{chate2020dry}
Chat\'{e} H.
\newblock Dry aligning dilute active matter.
\newblock {\em Annual Review of Condensed Matter Physics\/} {\bf 11} (2020)
  189--212.

\bibitem[{Romanczuk et~al.(2012)Romanczuk, B{\"a}r, Ebeling, Lindner, and
  Schimansky-Geier}]{romanczuk2012active}
Romanczuk P, B{\"a}r M, Ebeling W, Lindner B, Schimansky-Geier L.
\newblock Active {B}rownian particles.
\newblock {\em The European Physical Journal Special Topics\/} {\bf 202} (2012)
  1--162.

\bibitem[{Elgeti et~al.(2015)Elgeti, Winkler, and Gompper}]{elgeti2015physics}
Elgeti J, Winkler RG, Gompper G.
\newblock Physics of microswimmers~--~single particle motion and collective
  behavior: a review.
\newblock {\em Reports on Progress in Physics\/} {\bf 78} (2015) 056601.

\bibitem[{Keshavarz~Motamed and Maftoon(2021)}]{keshavarz2021systematic}
Keshavarz~Motamed P, Maftoon N.
\newblock A systematic approach for developing mechanistic models for realistic
  simulation of cancer cell motion and deformation.
\newblock {\em Scientific Reports\/} {\bf 11} (2021) 1--18.

\bibitem[{Nagel et~al.(2014)Nagel, Guven, Theves, Driscoll, Losert, and
  Beta}]{nagel_geometry-driven_2014}
Nagel O, Guven C, Theves M, Driscoll M, Losert W, Beta C.
\newblock Geometry-driven polarity in motile amoeboid cells {\bf 9} (2014)
  e113382.

\bibitem[{Theurkauff et~al.(2012)Theurkauff, Cottin-Bizonne, Palacci, Ybert,
  and Bocquet}]{theurkauff2012dynamic}
Theurkauff I, Cottin-Bizonne C, Palacci J, Ybert C, Bocquet L.
\newblock Dynamic clustering in active colloidal suspensions with chemical
  signaling.
\newblock {\em Physical Review Letters\/} {\bf 108} (2012) 268303.

\bibitem[{Huber et~al.(2018)Huber, Suzuki, Kr{\"u}ger, Frey, and
  Bausch}]{huber2018emergence}
Huber L, Suzuki R, Kr{\"u}ger T, Frey E, Bausch AR.
\newblock Emergence of coexisting ordered states in active matter systems.
\newblock {\em Science\/} {\bf 361} (2018) 255--258.

\bibitem[{Rappel(2016)}]{rappel2016cell}
Rappel WJ.
\newblock Cell--cell communication during collective migration.
\newblock {\em Proceedings of the National Academy of Sciences\/} {\bf 113}
  (2016) 1471--1473.

\bibitem[{Zimmermann et~al.(2016)Zimmermann, Camley, Rappel, and
  Levine}]{zimmermann2016contact}
Zimmermann J, Camley BA, Rappel WJ, Levine H.
\newblock Contact inhibition of locomotion determines cell--cell and
  cell--substrate forces in tissues.
\newblock {\em Proceedings of the National Academy of Sciences\/} {\bf 113}
  (2016) 2660--2665.

\bibitem[{van Haastert(2021)}]{van2021short}
van Haastert PJ.
\newblock Short-and long-term memory of moving amoeboid cells.
\newblock {\em PLoS ONE\/} {\bf 16} (2021) e0246345.

\bibitem[{Bosgraaf and Van~Haastert(2009)}]{bosgraaf2009ordered}
Bosgraaf L, Van~Haastert PJ.
\newblock The ordered extension of pseudopodia by amoeboid cells in the absence
  of external cues.
\newblock {\em PLoS ONE\/} {\bf 4} (2009) e5253.

\bibitem[{Andrew and Insall(2007)}]{andrew2007chemotaxis}
Andrew N, Insall RH.
\newblock Chemotaxis in shallow gradients is mediated independently of
  {P}td{I}ns 3-kinase by biased choices between random protrusions.
\newblock {\em Nature Cell Biology\/} {\bf 9} (2007) 193--200.

\bibitem[{Deforet et~al.(2014)Deforet, Hakim, Yevick, Duclos, and
  Silberzan}]{deforet2014emergence}
Deforet M, Hakim V, Yevick H, Duclos G, Silberzan P.
\newblock Emergence of collective modes and tri-dimensional structures from
  epithelial confinement.
\newblock {\em Nature Communications\/} {\bf 5} (2014) 3747.

\bibitem[{Doxzen et~al.(2013)Doxzen, Vedula, Leong, Hirata, Gov, Kabla
  et~al.}]{doxzen2013guidance}
Doxzen K, Vedula SRK, Leong MC, Hirata H, Gov NS, Kabla AJ, et~al.
\newblock Guidance of collective cell migration by substrate geometry.
\newblock {\em Integrative Biology\/} {\bf 5} (2013) 1026--1035.

\end{thebibliography}


\newpage
\clearpage

\section*{Figures and captions}


\begin{table}[]
\caption{Parameter values for the numerical model. The parameter $m_{rep}$, parametrizing the repulsive interactions, is introduced in this work and, therefore, no reference is given. }\label{table1}
\vspace{0.2cm}
	\begin{tabular}{@{}*{5}{l}}
		parameter  & value   & units   & meaning  & reference\\
		$D$        &   0.5  & $\mu m^2 /s$  &  diffusion coefficient & \cite{alonso2018modeling}\\
		$k_a$      &   1-6   & $s^{-1}$  &  reaction rate   & \cite{alonso2018modeling} \\
		$\rho$     &   0.02   & $s^{-1}$  &  degradation rate	 	& \cite{alonso2018modeling}\\
		$\sigma$   &   0.15    &  $s^{-2}$ &  noise strength	      & \cite{alonso2018modeling}\\
		$\tau$     &   2   & $pN s \mu m^{-2} $ & membrane dynamics time-scale  & \cite{shao2010computational}\\
		$\gamma$   &   2   & $pN$ & surface tension	& \cite{shao2010computational}     	\\
		$\epsilon$ &   0.75   & $\mu m$ & membrane thickness   & \cite{shao2010computational}\\
		$\beta$    &   22.22   & $pN \mu m^{-3} $ & parameter for total area constraint & \cite{alonso2018modeling} \\
		$A_0$ 	   &   113   & $\mu m^{2}$ &   area of the cell & \cite{alonso2018modeling}  \\
		$\delta_0$ & 0.5   & - &   bistability critical parameter & \cite{alonso2018modeling}\\
		$M$ 	   & 0.045 & $\mu m^{-2}$ &   strength of the global feedback input & \cite{alonso2018modeling}\\
		$k_{\eta}$ & 0.1   &  $s^{-1}$   &   Ornstein-Uhlembeck rate & \cite{alonso2018modeling}\\
		$\alpha$   & 3 & $pN \mu m^{-1}$ & active tension  & \cite{ziebert2012model}\\
		$Co$       & 28 & $\mu m^{2}$ & maximum area coverage by $c$ & \cite{alonso2018modeling} \\	
		$m_{rep}$   & 10 & $pN \mu m^{-1}$ & strength of cell-cell repulsion &  \\
	\end{tabular}
\end{table}

\end{document}